\documentclass[twocolumn]{aastex61}
\received{}
\revised{}
\accepted{ for publication in ApJ}
\submitjournal{ApJ}

\newcommand{\thestar}{J0949$-$1617}

\shorttitle{The first bonafide CEMP-$r+s$ star}
\shortauthors{Gull et al.}

\begin{document}
\title{The R-Process Alliance: Discovery of the first metal-poor star with a combined $r$- and $s$-process element signature\footnote{This paper includes data gathered with the 6.5\,m Magellan Telescopes located at Las Campanas Observatory, Chile. 
Based on observations made with ESO Telescopes at the La Silla Observatory under programme ID 095.D-0202 (PI: Placco).
Based on observations obtained at the Southern Astrophysical Research (SOAR) telescope (2016A-0019 - PI: Santucci), which is a joint project of the Minist\'{e}rio da Ci\^{e}ncia, Tecnologia, Inova\c{c}\~{a}os e Comunica\c{c}\~{a}oes (MCTIC) do Brasil, the U.S. National Optical Astronomy Observatory (NOAO), the University of North Carolina at Chapel Hill (UNC), and Michigan State University (MSU).
}}

\correspondingauthor{Anna Frebel}
\email{afrebel@mit.edu}

\author{Maude Gull}
\affiliation{Department of Physics \& Kavli Institute for Astrophysics and Space Research, Massachusetts Institute of Technology, Cambridge, MA 02139, USA}

\author{Anna Frebel}
\affiliation{Department of Physics \& Kavli Institute for Astrophysics and Space Research, Massachusetts Institute of Technology, Cambridge, MA 02139, USA}
\affiliation{Joint Institute for Nuclear Astrophysics - Center for Evolution of the Elements, USA}

\author{Madelyn G. Cain}
\affiliation{Department of Physics \& Kavli Institute for Astrophysics and Space Research, Massachusetts Institute of Technology, Cambridge, MA 02139, USA}

\author{Vinicius M. Placco}
\affiliation{Department of Physics, University of Notre Dame, Notre Dame, IN 46556, USA}
\affiliation{Joint Institute for Nuclear Astrophysics - Center for Evolution of the Elements, USA}

\author{Alexander P. Ji}
\affiliation{Department of Physics \& Kavli Institute for Astrophysics and Space Research, Massachusetts Institute of Technology, Cambridge, MA 02139, USA}
\affiliation{The Observatories of the Carnegie Institution of Washington, 813 Santa Barbara St., Pasadena, CA 91101, USA}
\affiliation{Hubble Fellow}

\author{Carlo Abate}
\affiliation{Argelander-Institut f\"ur Astronomie, 53121 Bonn, Germany}

\author{Rana Ezzeddine}
\affiliation{Joint Institute for Nuclear Astrophysics - Center for Evolution of the Elements, USA}
\affiliation{Department of Physics \& Kavli Institute for Astrophysics and Space Research, Massachusetts Institute of Technology, Cambridge, MA 02139, USA}

\author{Amanda I. Karakas}
\affiliation{Monash Centre for Astrophysics, School of Physics \& Astronomy, Monash University, Clayton, Melbourne 3800, Australia}

\author{Terese T. Hansen}
\affiliation{The Observatories of the Carnegie Institution of Washington, 813 Santa Barbara St., Pasadena, CA 91101, USA}

\author{Charli Sakari}
\affiliation{Department of Astronomy, University of Washington, Seattle, WA 98195-1580, USA}

\author{Erika M. Holmbeck}
\affiliation{Department of Physics, University of Notre Dame, Notre Dame, IN 46556, USA}
\affiliation{Joint Institute for Nuclear Astrophysics - Center for Evolution of the Elements, USA}

\author{Rafael M.\ Santucci}
\affiliation{Instituto de Estudos S\'ocio-Ambientais, Planet\'ario, 
Universidade Federal de Goi\'as, Goi\^ania, GO 74055-140, Brazil}
\affiliation{Instituto de F\'isica, Universidade Federal de Goi\'as, Campus
Samambaia, Goi\^ania, GO 74001-970, Brazil}

\author{Andrew R. Casey}
\affiliation{Monash Centre for Astrophysics, School of Physics \& Astronomy, Monash University, Clayton, Melbourne 3800, Australia}

\author{Timothy C. Beers}
\affiliation{Department of Physics, University of Notre Dame, Notre Dame, IN 46556, USA}
\affiliation{Joint Institute for Nuclear Astrophysics - Center for Evolution of the Elements,  USA}

\begin{abstract}

We present a high-resolution ($R\sim35,000$), high signal-to-noise (S/N$>200$) Magellan/MIKE spectrum of the star RAVE~J094921.8$-$161722, a bright ($V=11.3$) metal-poor red giant star with $\mbox{[Fe/H]} = -2.2$, 
identified as a carbon-enhanced metal-poor (CEMP) star from the RAVE survey. We report its detailed chemical abundance signature of light fusion elements and heavy neutron-capture elements. We find J0949$-$1617 to be a CEMP star with $s$-process enhancement that must have formed from gas enriched by a prior $r$-process event. Light neutron-capture elements follow a low-metallicity $s$-process pattern, while the heavier neutron-capture elements above Eu follow an $r$-process pattern. The Pb abundance is high, in line with an $s$-process origin. Thorium is also detected, as expected from an $r$-process origin, as Th is not produced in the $s$-process. We employ nucleosynthesis model predictions that take an initial $r$-process enhancement into account, and then determine the
mass transfer of carbon and $s$-process material from a putative more massive companion onto the observed star. The resulting abundances agree well with the observed pattern. We conclude that J0949$-$1617 is the first bonafide CEMP-$r+s$ star identified. This class of objects has previously been suggested to explain stars with neutron-capture element patterns that originate from neither the $r$- or $s$-process alone. We speculate that J0949$-$1617 formed in an environment similar to those of ultra-faint dwarf galaxies like Tucana\,III and Reticulum\,II, which were enriched in $r$-process elements by one or multiple neutron star mergers at the earliest times.

\end{abstract}


\section{Introduction}\label{s:intro}
The chemical abundances of metal-poor stars contain unique information about element nucleosynthesis in the early universe and the beginning of chemical evolution. While the production of light elements  (with Z $<$ 30) through fusion processes in the cores of stars and supernova explosions is relatively well-understood (e.g., \citealt{nomoto2006,Heger10}), there remain fundamental open questions regarding the production of the heavier elements beyond the iron-peak. Over the past few decades, studies of extremely metal-poor stars with $\mbox{[Fe/H]}<-3.0$ and also very metal-poor stars with $\mbox{[Fe/H]}<-2.0$ 
have provided critical insights into heavy element formation in the early universe \citep{ARAA, Frebel15}. Principally, two paths of neutron-capture onto seed nuclei are distinguished: the slow ($s$-) process and the rapid ($r$-) process. They each form about half of the isotopes that constitute all the heavy-elements known from the periodic table. These processes operate in very different astrophysical sites. The $s$-process elements are produced under the H-burning shell in evolved thermally-pulsing asymptotic giant branch (AGB) stars (e.g., \citealt{gallino1998,Karakas10,Lugaro12}). A relatively low neutron flux is required, as elements are built-up during multiple thermal pulses over a time span of 10,000 years. The $r$-process requires a much higher neutron flux, and occurs within 1-2 seconds (e.g., \citealt{goriely11,Korobkin12}). Recent results suggest neutron star mergers to be the primary source of the entire range of $r$-process elements \citep{ishimaru15,Ji16c,Ji16b}. Core-collapse supernovae may still provide lighter neutron-capture elements in smaller quantities (e.g., \citealt{izutani,Arcones11,Hansen12}).

Knowledge about nucleosynthesis processes that were in operation at early times, acquired from theoretical studies and nuclear physics experiments, has, step-by-step, been validated by observations of metal-poor stars with particular chemical signatures. The so-called CEMP-$s$ stars are ordinary very metal-poor halo stars that show the signature of $s$-process together with carbon in their spectrum. The carbon and $s$-process material was provided to the star by a binary companion that went through its AGB phase during which these elements were created. Many comparisons of theoretical predictions and observed abundance signatures have confirmed this picture \citep{bisterzo09,Lugaro12,Abate2015-2}. Radial velocity variations of many of these stars have provided additional validation that these stars are orbiting a now unseen companion \citep{lucatello2005,Starkenburg2014, hansen2016a}. 
Similarly, many extremely metal-poor stars show the clear signature of the $r$-process in their spectrum (e.g., \citealt{Sneden08}). The resulting $r$-process pattern matches that of the scaled solar $r$-process component. Most of these stars ($\sim80\%$) are not part of a binary system \citep{Hansen15}; their observed binary fractions ($\sim$18\%) are consistent with that of other metal-poor giants in the halo ($\sim$16\%; \citealt{carney2003}).  Thus, there is no evidence that their $r$-process enhancement is causally linked to the binary nature of the stars.
Rather, these stars were born from gas that was previously enriched in $r$-process elements. A handful of $r$-process-enhanced stars are also found among the CEMP stars, the so-called CEMP-$r$ stars (\citealt{ARAA}), including the canonical highly $r$-process-enhanced star CS~22892-052.

There have also been about two dozen CEMP stars found to date with large enhancements in neutron-capture elements that fit neither an $s$-process pattern nor that of an $r$-process. \citet{ARAA} introduced the notation \textquotedblleft CEMP-$r/s$" to describe such stars, a choice that intentionally remained silent on their specific origin, which was unclear at the time.
Over the years, multiple attempts involving various models and scenarios to explain the observed abundances of CEMP-$r/s$ stars with combined contributions from the two processes largely failed \citep{cohen2003, ivans05,jonsell06,Abate2016}. In addition, not all of these stars exhibit
a common, distinct pattern, making it very challenging to explain. 
However, recent progress in nucleosynthesis calculations suggests the existence of an intermediate neutron-capture process (the $i$-process, originally suggested by \citealt{cowan77}) which also operates in AGB 
stars, possibly those of higher mass than associated with CEMP-$s$ progenitors. Most of the stars previously categorized as CEMP-$r/s$ stars have been found to be $i$-process stars \citep{Hampel16,roederer2016}.  

The possible existence of a CEMP star that shows the combined signature of the $r$-process and an $s$-process remains viable, however. Such an object would have formed in a binary system formed from previously $r$-process enriched gas which later received material from a mass-transfer event involving a companion AGB star. Here we report the discovery the first bonafide CEMP-$r+s$ metal-poor star, RAVE~J094921.8$-$161722, which appears to display a combination of both the $s$-process and the $r$-process. Section 2 
describes the recognition of this star as a CEMP star, and summarizes the medium- and high-resolution observations.
Our abundance analysis is described in Section~3. The \textquotedblleft $r+s$" nature of the star is detailed in Section~4, and we consider the inferred old age of  RAVE~J094921.8$-$161722 in Section~5. Our conclusions are provided in Section~6.

\section{Observations, line measurements and stellar parameters}

RAVE~J094921.8$-$161722 (hereafter J0949$-$1617; with R.A. = 09:49:21.8, Dec. = $-$16:17:22.0; $V = 11.3$) was first identified as a very metal-poor star candidate from RAVE DR5 \citep{kunder17}, and
followed-up with medium-resolution spectroscopy obtained with the
ESO/NTT and the SOAR~4.1\,m telescopes.

\begin{figure*}[!ht]
 \begin{center}
  \includegraphics[clip=true,width=16cm]{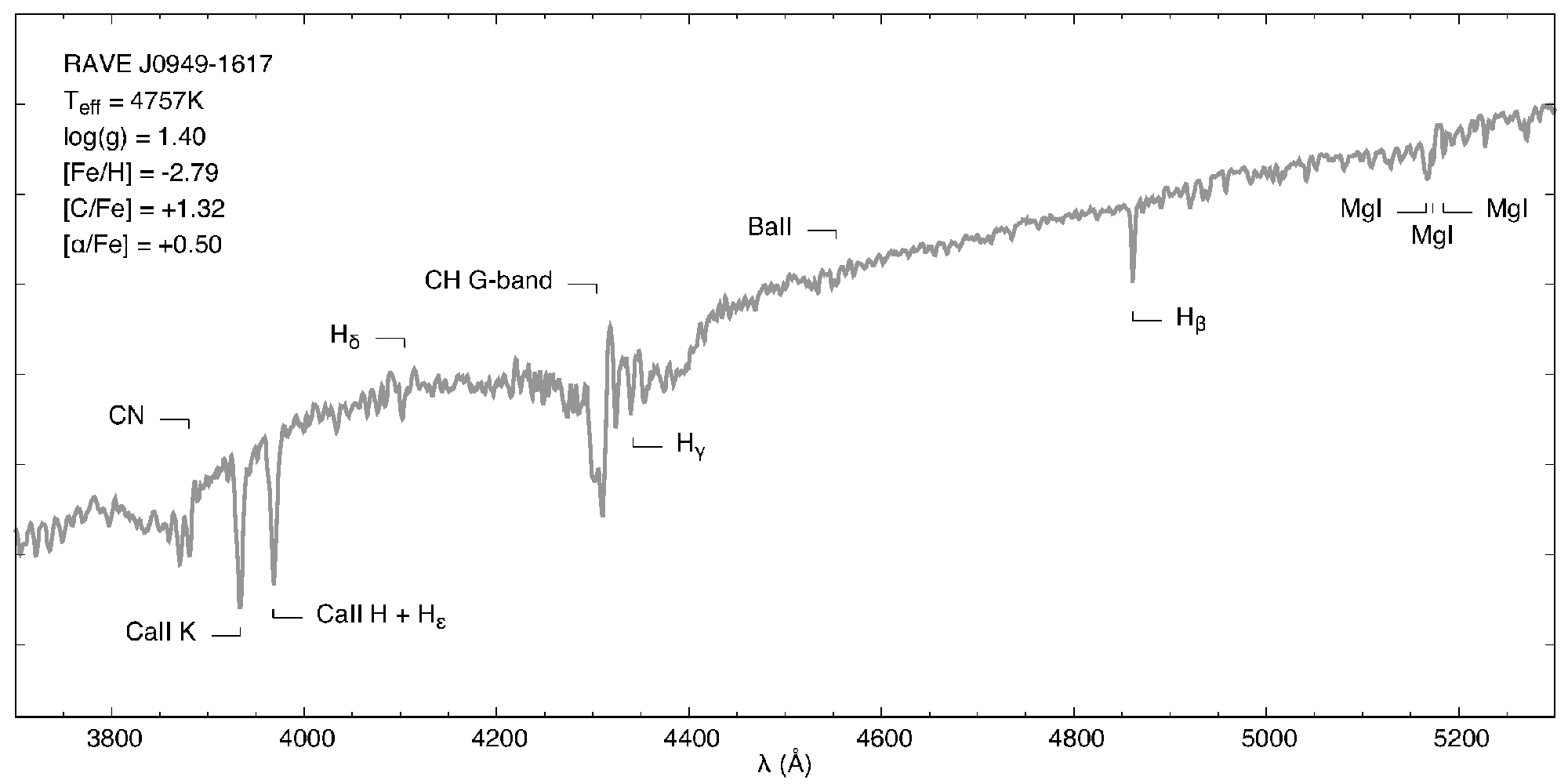} 
  \caption{\label{med} Medium-resolution ESO/NTT spectrum of \thestar. A strong CH $G$-band is readily identifiable, leading to the recognition of \thestar\ as a CEMP star.  
  }
 \end{center}
\end{figure*}

The NTT/EFOSC-2 data were gathered in 2015A, and employed Grism\#7
(600\,gr\,mm$^{\rm{-1}}$) and a 1$\farcs$0 slit, covering the wavelength range 3300-5100\,{\AA}. This combination yielded a resolving power $R\sim 1,900$ and signal-to-noise ratio S/N$\sim 40$ per pixel at 3900\,{\AA}.  The SOAR observations were carried out in 2016A, using the Goodman Spectrograph. The observing setup was the 600\,l\,mm$^{\rm{-1}}$ grating, the blue setting, and a 1$\farcs$03 slit, covering the wavelength range 3550-5500\,{\AA}, yielding a resolving power of $R\sim 1,500$ and S/N$\sim 50$ per pixel at 3900\,{\AA}. The
calibration frames in both cases included HgAr and Cu arc lamp exposures (taken following the science observations), bias frames, and quartz-lamp flatfields. Calibration and extraction were performed using standard IRAF\footnote{\href{http://iraf.noao.edu} {http://iraf.noao.edu}.} packages.
Stellar atmospheric parameters for J0949$-$1617 were determined from the medium-resolution spectra, along with 2MASS $J-K$ colors \citep{2MASS}, using the n-SSPP pipeline \citep{beers14}. The estimates obtained were: T$_{\rm eff} = 4757$\,K, $\log g = 1.4$, $\mbox{[Fe/H]} = -$2.8, confirming its status as a very metal-poor star.  
Note that the RAVE spectra only cover the region of the Ca\,I triplet, but as can be seen in Figure~\ref{med}, our medium-resolution spectra cover
the CH $G$-band, enabling a measurement of $\mbox{[C/Fe]} = +1.3$ (includes a +0.4\,dex correction for the effects of stellar evolution from \citealt{PLACCO14}). 
This satisfies the definition of a CEMP star, usually taken to be $\mbox{[C/Fe]}> + 0.7$. We also determined an estimate of  $\mbox{[$\alpha$/Fe]} = + 0.5$ from the medium-resolution spectra.

\thestar\ was then observed with the MIKE spectrograph \citep{Bernstein03} on the Magellan-Clay telescope at Las Campanas Observatory on April 15 and 16, 2016, and again on June 5, 2017. Conditions were excellent during the 2017 run, with seeing of 0\farcs5. We thus opted to repeat our initial analysis using only the 2017 spectrum. The 0\farcs7 slit employed yields a nominal spectral resolving power of $R\sim$28,000 in the red and $R\sim$35,000 in the blue wavelength regime, but the excellent seeing increased the resolving power to about 40,000 to 45,000 in the blue. The total exposure time was 30\,min in 2016 and 50\,min in 2017. 
Data reductions were carried out with the MIKE Carnegie Python pipeline \citep{Kelson03}. The resulting S/N per pixel in the 2017 spectrum is 190 at $\sim$4000\,{\AA}, 370 at $\sim$4700\,{\AA}, 280 at $\sim$5200 \AA, and 470 at $\sim$6000\,{\AA}. 

We have complied heliocentric radial-velocity measurements for \thestar\ covering $\sim$13 years, although with a seven year gap from 2009 to 2016; these are listed in Table~\ref{rv}.  
Typical uncertainties of the LCO/Magellan measurements are 1-2\,km\,s$^{-1}$, from comparison with well-studied stars observed during the same night. 
Repeat RAVE measurements taken from DR5 \citep{kunder17} have uncertainties of 0.6-0.7\,km\,s$^{-1}$. 
High-resolution follow-up spectra obtained with the Astrophysical Research Consortium Echelle Spectrograph on the 3.5\,m telescope at Apache Point Observatory yield velocity uncertainties $\sim2$\,km\,s$^{-1}$.  
Observations taken with the South African Large Telescope using the high-resolution spectrograph have $\pm 1$\,km\,s$^{-1}$ uncertainties. All measurements agree remarkable well each other, which strongly constrains potential radial-velocity variations of \thestar\ to less than a few km\,s$^{-1}$. This is unlike what is expected for most $s$-process stars, which typical exhibit clear radial-velocity variations due to their binarity. However, as shown in \citet{hansen2016a}, only $\sim80\%$ of their $s$-process star sample exhibited clear radial-velocity variations. \thestar\ could resemble the remaining $\sim20\%$, perhaps having an orbital motion that is simply not detectable due to the system's orientation being face-on with respect to our line-of-sight, or possibly because the orbit is very wide ($P_{\mathsf{orb}}>5000$ days or so), or perhaps because carbon and $s$-process material was added to its natal gas cloud in large amounts prior to the formation of \thestar. For reference, the $r$-process star HE~1523$-$0901 has variations of only 0.3\,km\,s$^{-1}$ \citep{Hansen15}; \thestar\ could easily show similar behavior. Future radial-velocity monitoring would clearly be helpful for assessing this issue. For the remainder of the paper, however, we assume that \thestar\ is a member of a binary system, and model its abundance pattern accordingly. Regardless of whether \thestar\ is in a binary system, its large radial velocity suggests the star to be a member of the metal-poor outer-halo population(\citealt{carollo}, \citealt{carollo10}, \citealt{beers2012}).

\begin{deluxetable}{lrc}
\tablecolumns{6}
\tablecaption{\label{rv} Radial Velocities of the CEMP-$r+s$ Metal-Poor Star \thestar}
\tablehead{ 
\colhead{UT date} &
\colhead{$v_{\rm helio}$} & 
\colhead{Observatory/} \\
\colhead{}&\colhead{[km\,s$^{-1}$]}&\colhead{Survey}}
\startdata
2004 April 08 & 391.2&  RAVE \\
2006 April 21 & 391.4&  RAVE \\
2009 March 03 & 391.7&  RAVE \\
2009 March 04 & 390.7&  RAVE \\
2009 March 18 & 391.3&  RAVE \\
2009 May 25   & 390.1&  RAVE \\
2016 January 17 & 391.2 & APO\\
2016 January 28 & 390.1 & APO\\
2016 February 1 & 390.1 & SALT \\
2016 April 15   & 389.5 & LCO \\
2017 March 7 & 390.3 & APO \\
2017 June 5 & 392.5 & LCO \\
\enddata
\tablecomments{RAVE: Radial Velocity Experiment, APO: Apache Point Observatory, SALT: South African Large Telescope, LCO: Las Campanas Observatory.}
\end{deluxetable}

After shifting the spectrum to rest wavelengths, we measured equivalent widths of various absorption lines, including 172 Fe\,I and 23 Fe\,II lines, by fitting Gaussian profiles to them. The equivalent widths are presented in Table~1. In the process of measuring equivalent widths, we noticed that numerous CH lines throughout the spectrum resulted in severe blending of many usually clean absorption lines. We thus resorted to eliminating all Fe lines leading up to the CH bandhead at 4313\,{\AA}. Furthermore, we discarded all lines between 5100\,{\AA} and 5160\,{\AA}, i.e., leading up to the C$_{2}$ bandhead. 

We employ a 1D plane-parallel model atmosphere with $\alpha$-enhancement \citep{Castelli04} and the 2014 version of the MOOG analysis code \citep{Sneden73}, to which we added Rayleigh scattering (following \citealt{Sobeck11}).
All of this is integrated into an updated version of a custom-made analysis tool first described in \citet{Casey14}. We compute elemental line abundances assuming local thermodynamic equilibrium (LTE). We then determined the stellar parameters spectroscopically, following the procedure outlined in \citet{Frebel13}. We obtain an effective temperature of T$_{eff} = 4855$\,K, surface gravity of $\log g = 1.60$, microturbulence v$_{micr}= 1.90$\,km\,s$^{-1}$, and metallicity $\mbox{[Fe/H]} = -2.22$. Placing the star on a 12\,Gyr old theoretical isochrone \citep{Y2_iso} shows good agreement.

We also obtain spectroscopic stellar parameters by calculating individual Fe-line abundance correction assuming non-LTE. This is based on the quantum-fitting method further described in \citet{ezzeddine16a}, and applied to a sample of the most iron-poor stars in \citet{ezzedine17a}. Our non-LTE results are as follows:
T$_{eff} = 4750$\,K, $\log g = 2.1$, v$_{micr}= 1.80$\,km\,s$^{-1}$, and $\mbox{[Fe/H]} = -2.08$. While temperature is lower in the non-LTE case, the metallicity is increased. From forcing an ionization balance, the surface gravity is correspondingly higher as well, since Fe\,I is primarily affected by non-LTE. Following \citet{ezzedine17a}, the increase of 0.14\,dex in [Fe/H] agrees well with an increase of 0.16\,dex derived from 
\begin{equation} \label{correction_fit} \Delta \mbox{[Fe/H]} = -0.14\, \mathrm{[Fe/H]}_{\mathrm{LTE}} - 0.15
\end{equation}
Also, the scatter in the relations (abundance vs. excitation potential and reduced equivalent width) used to determined the stellar parameters is reduced from 0.13 to 0.10\,dex for Fe\,I-line and from 0.10 to 0.06\,dex for Fe\,II-line abundances. This behavior was found in \citet{ezzedine17a} for the most iron-poor stars, and appears to apply to mildly metal-poor stars like \thestar\ as well. In the following, however, we adopt the LTE abundances to produce consistent abundance ratios that can  be readily compared to literature values. 

We estimate uncertainties in the stellar parameters as follows. From varying the slope of the line abundances as a function of excitation potential of the lines within its uncertainty, we find $\sigma_{\rm Teff} = 80$\,K. The standard deviation of Fe\,I lines abundances is 0.13\,dex. We note that the resulting standard error would be 0.01\,dex, which is unrealistically small. We thus adopt the standard deviation as our Fe abundance uncertainty, as it is a  typical value for high-resolution, high S/N spectral analyses. Varying the Fe\,I abundance by 0.13\,dex results in changes in surface gravity of 0.27\,dex, which we adopt as the uncertainty in this parameter. Finally, we adopt an uncertainty for the microturbulence of 0.3\,km\,s$^{-1}$. We also determined stellar parameters without excluding lines in the CH and C$_{2}$ regions to investigate the potential impact. We found results within the uncertainties of our final values, but, as expected, the scatter among line abundances was much larger due to the effect of blending on many lines, especially for the blue lines. 

For spectral lines and blended features of other elements, we performed spectrum synthesis. During the analysis, it became clear that this star is not only cool, and very enhanced in carbon ($\mbox{[C/Fe]}=1.35$, see below) but also enriched in neutron-capture elements. We report details on individual chemical elements in Section 3 below.


\startlongtable
\begin{deluxetable}{lrrrrr}
\tablecolumns{6}
\tablecaption{\label{abund} Chemical Abundances of the  CEMP-$r+s$ Metal-Poor Star \thestar}
\tablehead{ 
\colhead{Species} &
\colhead{$\log\epsilon (\mbox{X})$} & 
\colhead{[X/H]} & 
\colhead{[X/Fe]}& 
\colhead{$N$}& 
\colhead{$\sigma$}}
\startdata
C (CH)  &   7.38 &  $-$1.05 &     1.17 &  2 &  0.10\\
 (CH)$_{\rm {corr}}$  &  \nodata &  $-$0.87 &   1.35 &  \nodata &  \nodata \\
N (NH) &    6.20 &  $-$1.63 &     0.59 &  1 &  0.20\\
O\,I   &    7.28 &  $-$1.41 &     0.81 &  2 &  0.10\\
Na\,I  &    4.30 &  $-$1.94 &     0.28 &  4 &  0.17\\
Mg\,I  &    5.79 &  $-$1.81 &     0.41 &  9 &  0.12\\
Al\,I  &    3.75 &  $-$2.70 &  $-$0.48 &  1 &  0.10 \\
Si\,I  &    5.51 &  $-$2.00 &     0.22 &  4 &  0.10\\
Ca\,I  &    4.58 &  $-$1.76 &     0.46 & 23 &  0.21\\
Sc\,II &    1.11 &  $-$2.04 &     0.18 & 15 &  0.18\\
Ti\,I  &    2.99 &  $-$1.96 &     0.26 & 32 &  0.17\\
Ti\,II &    3.20 &  $-$1.75 &     0.47 & 48 &  0.22\\
V\,II  &    1.81 &  $-$2.12 &  $-$0.08 &  4 &  0.10\\ 
Cr\,I  &    3.26 &  $-$2.38 &  $-$0.16 & 19 &  0.19\\
Cr\,II &    3.67 &  $-$2.11 &  $-$1.97 &  2 &  0.10\\
Mn\,I &    2.73 &  $-$2.70 &  $-$0.48 &  3 &  0.10\\
Fe\,I  &    5.28 &  $-$2.22 &     0.00 &172 &  0.13\\
Fe\,II &    5.30 &  $-$2.20 &     0.00 & 23 &  0.10\\
Co\,I &    2.74 &  $-$2.25 &  $-$0.03 &  6 &  0.19\\
Ni\,I  &    3.96 &  $-$2.41 &  $-$0.04 & 29 &  0.21\\
Zn\,I  &    2.59 &  $-$1.96 &     0.26 &  2 &  0.11\\
Sr\,II &    1.09 &  $-$1.78 &     0.44 &  3 &  0.14\\
Y\,II  &    0.37 &  $-$1.84 &     0.38 & 16 &  0.19\\
Zr\,II &    0.95 &  $-$1.63 &     0.59 & 21 &  0.18\\
Ru\,I  &    0.54 &  $-$1.21 &     1.01 &  4 &  0.15\\
Rh\,I  & $-$0.40 &  $-$1.31 &     0.91 &  1 & 0.20 \\
Pd\,I  & $-$0.04 &  $-$1.61 &     0.61 &  3 &  0.17\\
Ba\,II &    0.95 &  $-$1.23 &     0.99 &  6 &  0.12\\
La\,II & $-$0.05 &  $-$1.15 &     1.07 & 19 &  0.10\\
Ce\,II &    0.35 &  $-$1.23 &     0.99 & 30 &  0.19\\
Pr\,II & $-$0.59 &  $-$1.31 &     0.91 &  9 &  0.12\\
Nd\,II &    0.13 &  $-$1.29 &     0.93 & 46 &  0.12\\
Sm\,II & $-$0.41 &  $-$1.37 &     0.85 & 19 &  0.13\\
Eu\,II & $-$1.09 &  $-$1.61 &     0.61 &  4 &  0.10\\
Gd\,II & $-$0.46 &  $-$1.53 &     0.69 &  5 &  0.10\\
Tb\,II & $-$1.23 &  $-$1.53 &     0.69 &  4 &  0.18\\
Dy\,II & $-$0.35 &  $-$1.46 &     0.77 &  8 &  0.19\\
Ho\,II & $-$1.08 &  $-$1.56 &     0.66 &  6 &  0.21\\
Er\,II & $-$0.59 &  $-$1.51 &     0.71 &  4 &  0.11\\
Tm\,II & $-$1.40 &  $-$1.50 &     0.72 &  4 &  0.12\\
Lu\,II & $-$1.20 &  $-$1.30 &     0.92 &  1 &  0.20\\
Hf\,II & $-$0.65 &  $-$1.50 &     0.72 &  2 &  0.10\\
Os\,I  & $-$0.15 &  $-$1.55 &     0.67 &  2 &  0.12\\
Ir\,I  & $-$0.25 &  $-$1.63 &     0.59 &  1 &  0.30\\
Pb\,I  &    1.21 &  $-$0.54 &     1.68 &  1 &  0.30\\
Th\,II & $-$1.70 &  $-$1.72 &     0.50 &  1 &  0.20
\enddata
\tablecomments{
Stellar parameters for \thestar\ are T$_{eff} = 4855$\,K, $\log g = 1.60$, v$_{micr}= 1.90$\,km\,s$^{-1}$, and $\mbox{[Fe/H]} = -2.22$.  
[X/Fe] ratios are computed using the [Fe\,I/H] abundance and solar abundances from \citet{Asplund09}. $\sigma$ denotes the standard deviation of line abundances. For abundances measured from only one line, we adopt a nominal uncertainty of 0.1 - 0.30\,dex, depending on the quality of the fit. }
\end{deluxetable}

\section{Chemical Abundance Analysis}\label{s:analysis}

\subsection{Abundances up to Zinc} 
\begin{figure*}[!ht]
 \begin{center}
  \includegraphics[clip=true,width=19cm]{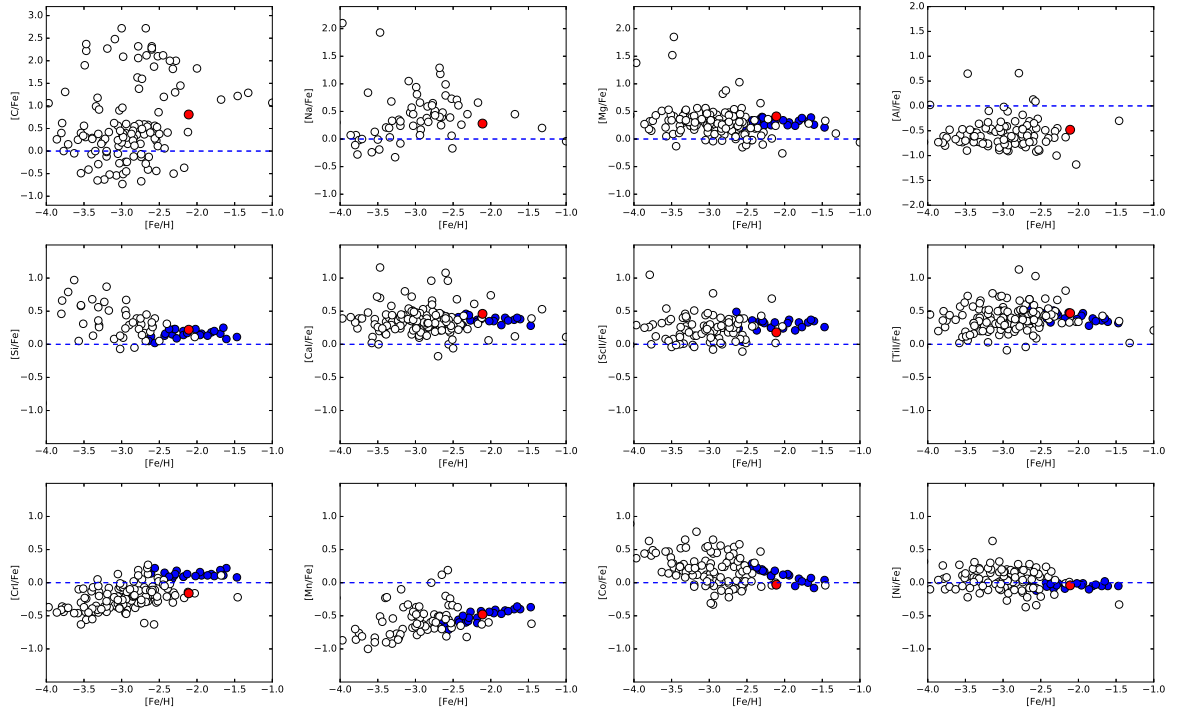} 
  \vspace{-4cm}
  \caption{\label{abundplot_light}Chemical abundances of various light elements in J0949$-$1617 (red circle). There is excellent agreement with the non-CEMP stars of \citet{yong13_II} (small black open circles) and \citet{reggi17} (small blue filled circles). }
 \end{center}
\end{figure*}

Using equivalent-width measurements, or spectrum synthesis when appropriate, we determined chemical abundances of 16 light elements up to zinc that are typically measured in metal-poor halo stars. Our final abundances for \thestar\ are listed in Table~1. Solar abundances of \citep{Asplund09} are used to calculate abundance ratios, [X/Fe].  Figure~\ref{abundplot_light} shows our abundance results, plotted along with the non-CEMP stars of \citet{yong13_II} and \citep{reggi17}, although we exclude Na and Al from \citep{reggi17} since their NLTE abundances do not compare with our LTE values.
For most elements, it appears that \thestar\ does not stand out from the majority of the halo stars with similar Fe abundances, which can be assumed to have formed from well-mixed gas enriched by core-collapse supernovae of previous stellar generations. 

One exception is carbon. \thestar\ is highly carbon enhanced, having $\mbox{[C/Fe]}= + 1.17$, as measured from the C$_{2}$ bandhead, using a linelist from \citet{Masseron14}. The CH $G$-band was saturated in our spectrum and not used. Considering that \thestar\ is an evolved red giant, and has thus undergone some level of internal mixing, we obtain a correction for the carbon abundance from \citet{PLACCO14} that account for the effect of decreasing carbon levels as stars ascend the giant branch. The correction is 
0.18\,dex, which brings the final C abundance to $\mbox{[C/Fe]}= + 1.35$.
We obtained a $^{12}$C/$^{13}$C of 19
ratio by fitting to a doublet of $^{12}$C and $^{13}$C lines in the 4217\,{\AA} region, as shown in Figure~\ref{abundplot_all}. The ratio obtained also fits the feature in the 4019\,{\AA} region. 
The ratio is in line with the star's evolutionary status. Together with enhancements in neutron-capture elements, in particular Ba, this suggests that 
J0949$-$1617 is a CEMP-$s$ star. Then, the carbon abundance does not reflect the abundances of the natal gas cloud, nor do the neutron-capture abundances. 
Instead, \thestar\ must have received 
its carbon and $s$-process elements during a mass-transfer event from a companion star that went through an AGB phase. As described further below, an $s$-process origin indeed partially explains the unique chemical signature of this star.

\subsection{Neutron-Capture Element Abundances from Strontium to Thorium} 

\begin{figure*}[!ht]
\begin{center}
\includegraphics[clip=true,width=8.5cm]{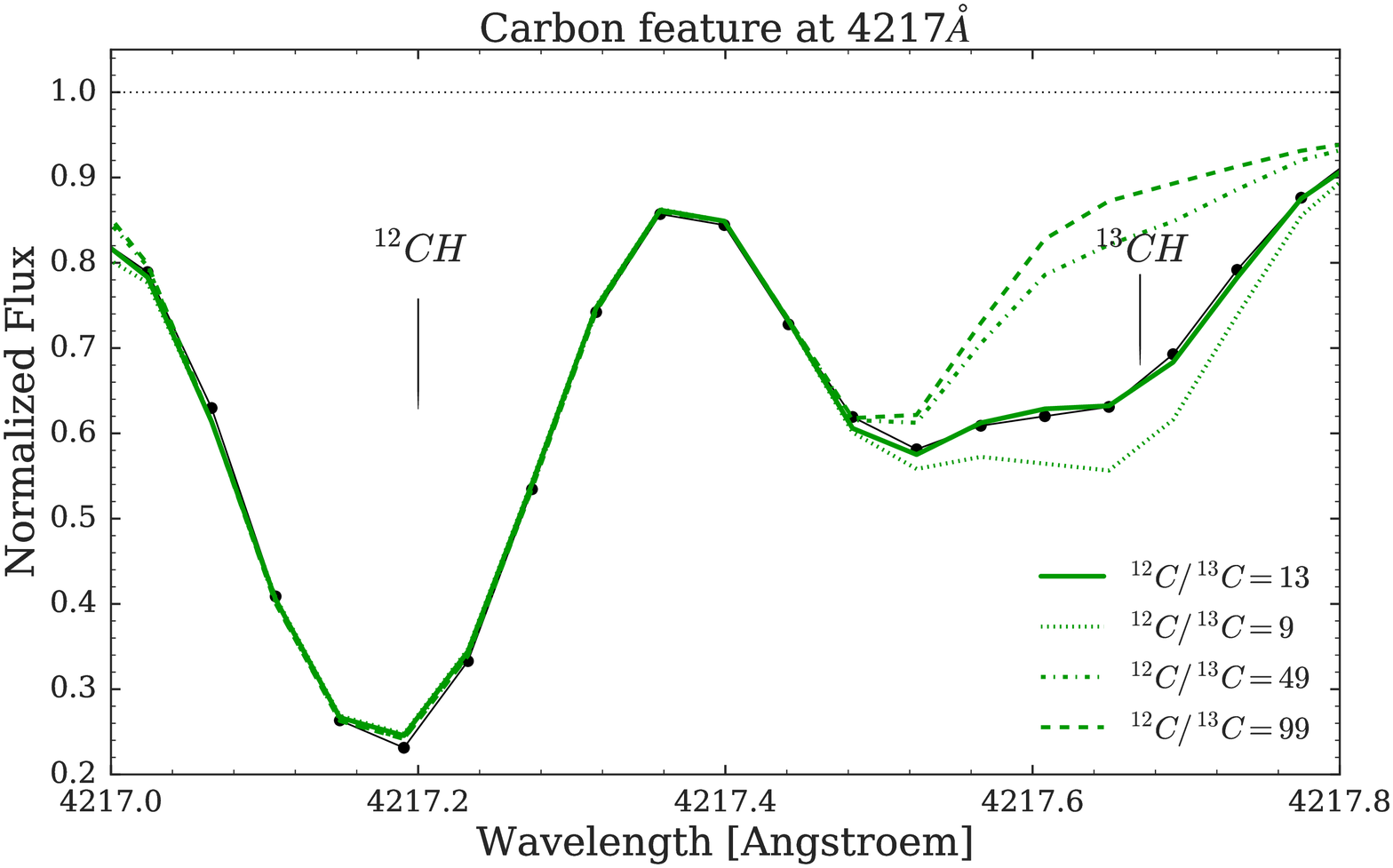}   
\includegraphics[clip=true,width=7.9cm]{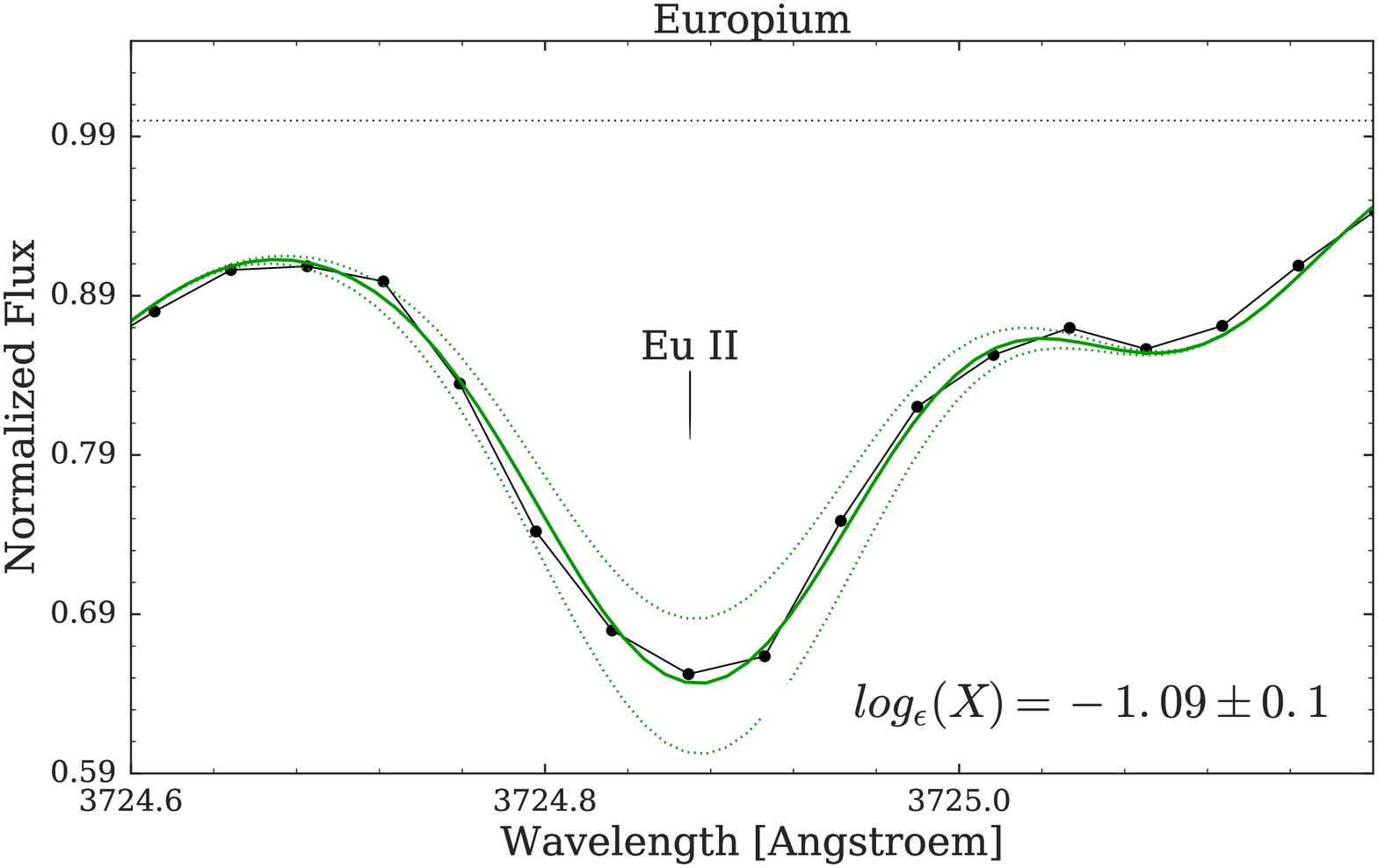}  \\
\includegraphics[clip=true,width=8.5cm]{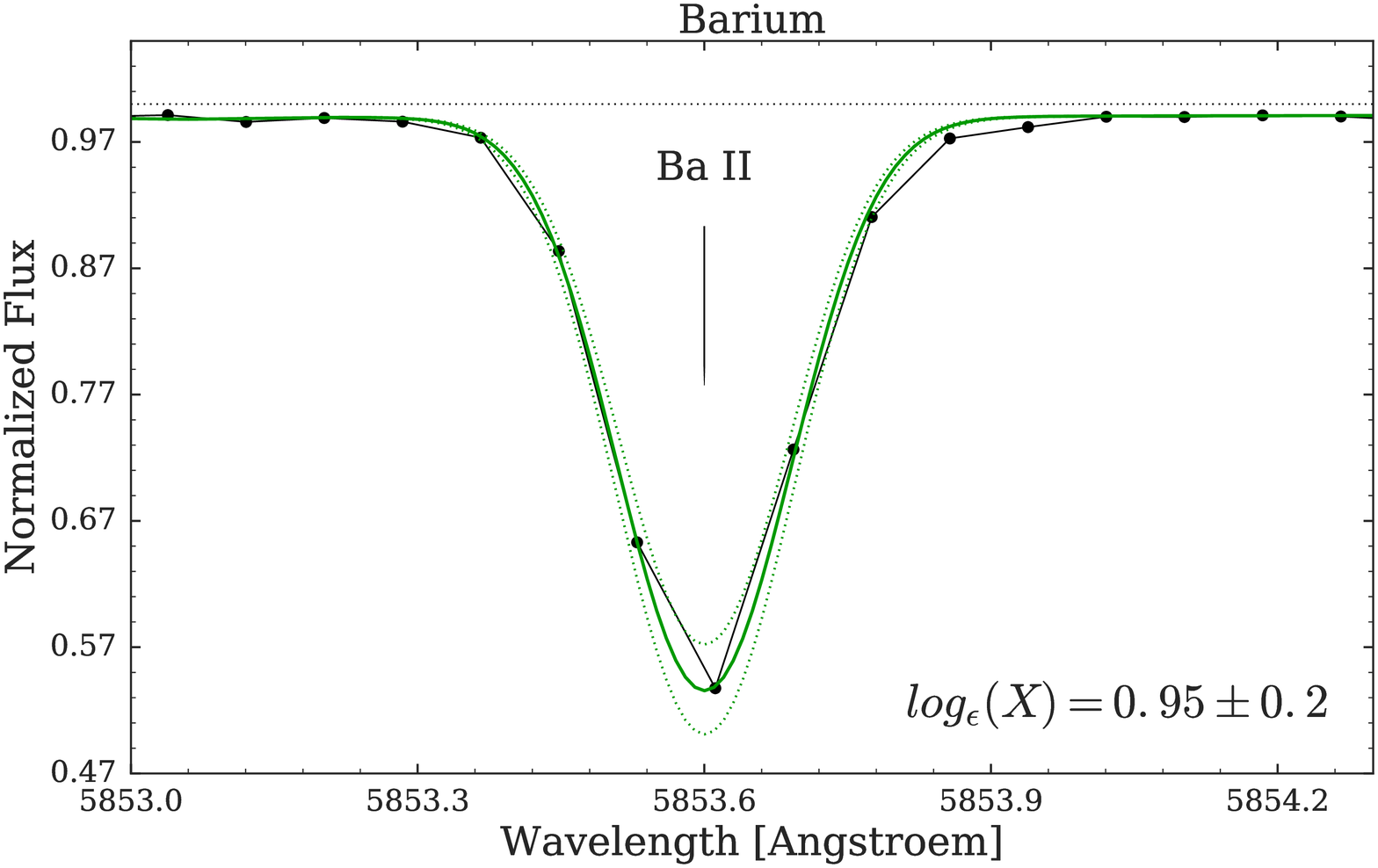} 
\includegraphics[clip=true,width=8.0cm]{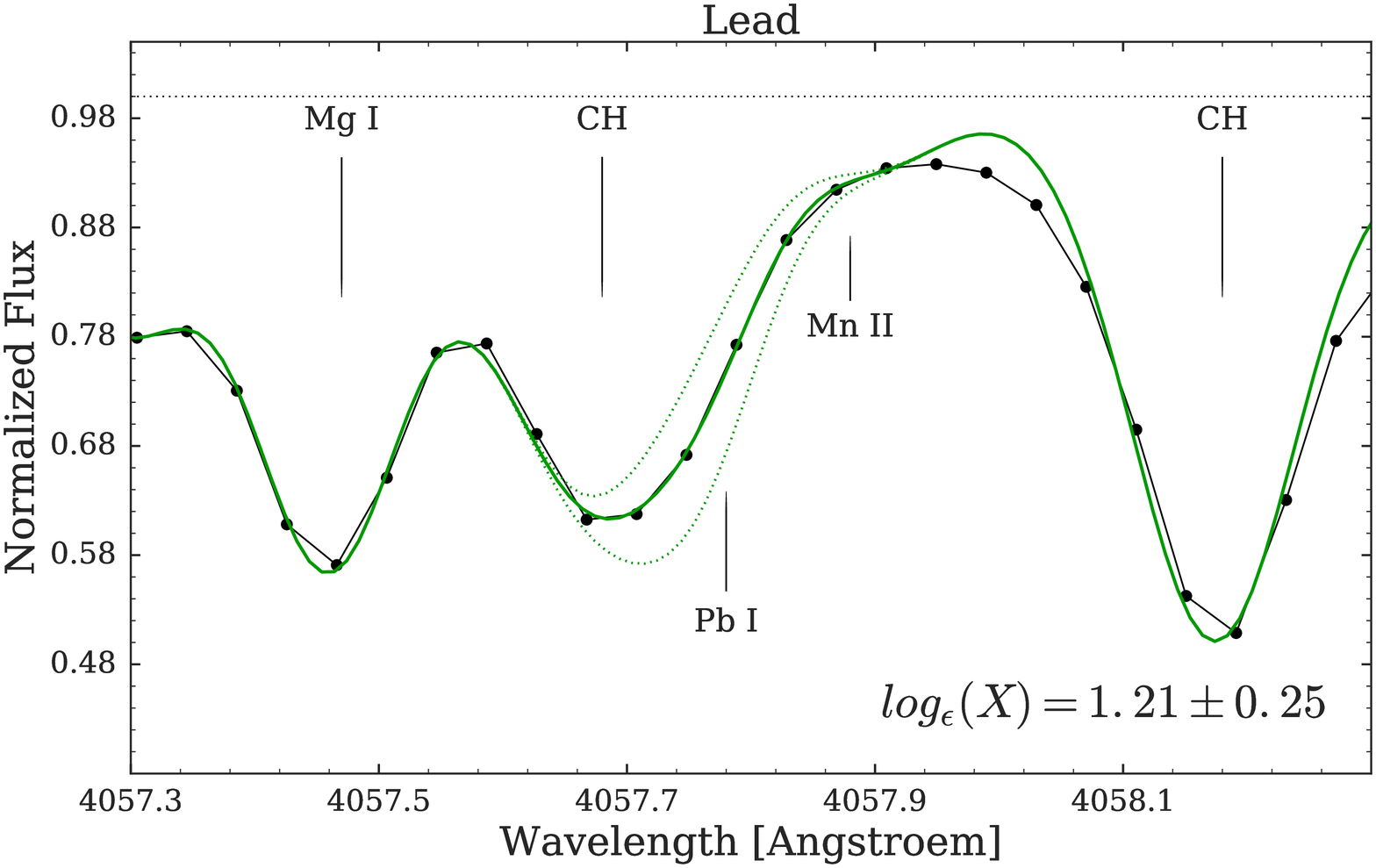} \\
\caption{\label{abundplot_all} Portions of the Magellan/MIKE spectrum of \thestar\ (shown as dashed lines) near the lines of CH isotopes at 4217\,{\AA} (top left), Ba\,II at 5853\,{\AA} (bottom left), Eu\,II at 3724\,{\AA} (top right), and Pb\,I at 4057\,{\AA} (bottom right). Best-fit synthetic spectra are also shown (green solid line) together with abundance variations (green dotted line) of $\pm$0.1\,dex (for Eu), $\pm$0.2\,dex (for Ba), and $\pm$0.25\,dex (for Pb). Different isotope ratios are shown for the CH features. Some prominent absorption lines are indicated.} 
\end{center}
\end{figure*}

\begin{figure*}[!ht]
\begin{center}
    \includegraphics[clip=true,width=19cm]{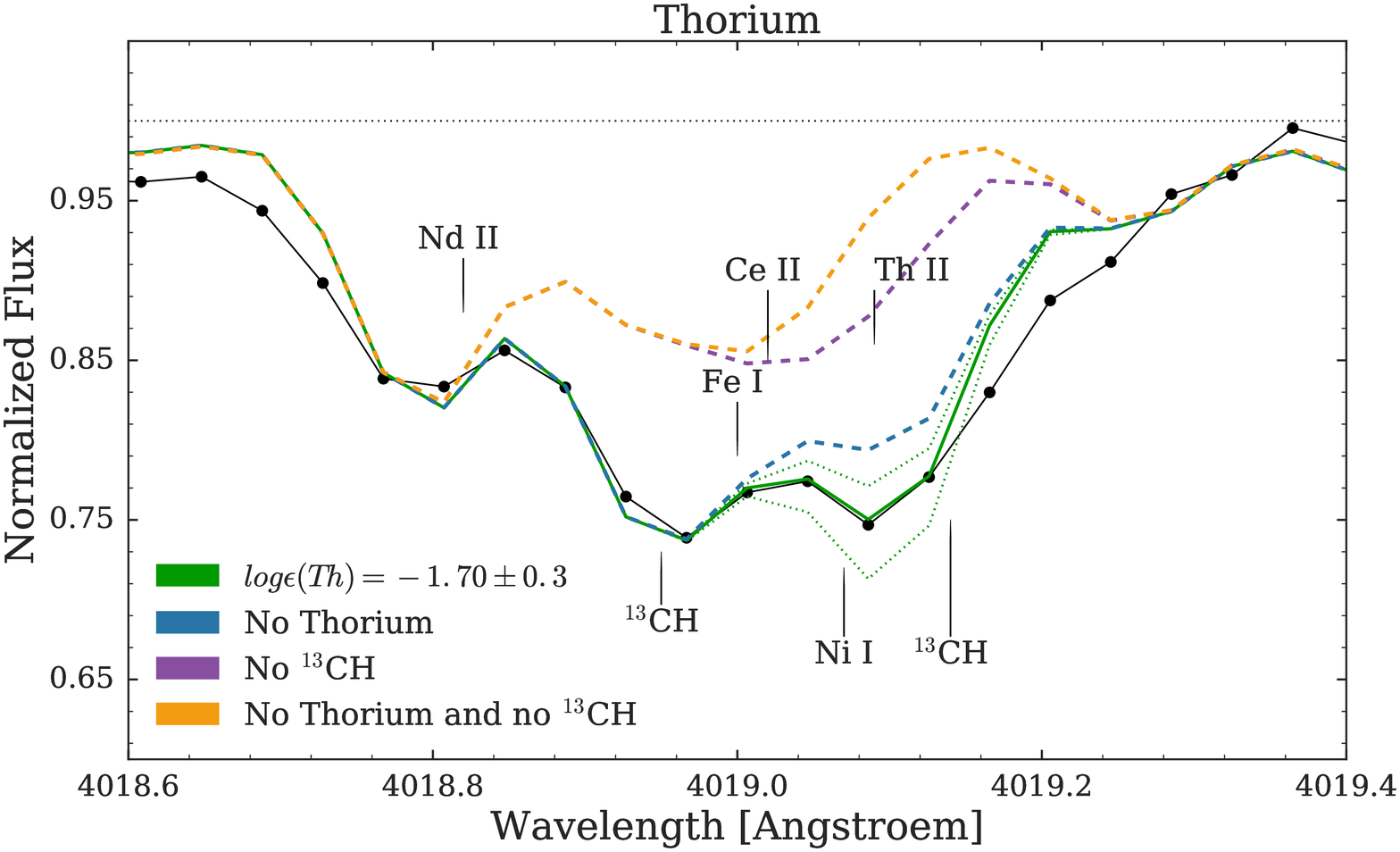}     
  \caption{\label{abundplot_Th} Portion of the Magellan/MIKE spectrum of \thestar\ (shown as dashed lines) near the Th\,II line at 4019\,{\AA} Synthetic spectra are included without any Th contributions (blue line), without any $^{13}$CH contributions (purple line) and  without either contribution (orange line). The line list used to synthesize this region is provided in Table \ref{synth_table}.} 
\end{center}
\end{figure*}

Using predominantly spectrum synthesis to account for blending and hyperfine structure of absorption features, we obtained chemical abundances for 25 neutron-capture elements between strontium and thorium. Our final abundances are listed in Table~1. Figures~\ref{abundplot_all} and \ref{abundplot_Th}  show examples of spectrum-synthesis abundance measurements for lines of CH, Ba, Eu, Pb, and Th. 

As for the thorium abundance, the $\lambda4019$ feature is blended with $^{13}$CH and other elemental lines. In Table~\ref{synth_table}, we provide the line list used for the synthesis of the region. We performed tests to ensure that the observed feature indeed includes a measurable thorium contribution, by e.g., trying to replicate the observed spectrum without any thorium present or attempting to fit the spectrum with just a $^{13}$CH contribution. However, it was not possible to adequately match the spectrum without producing discrepancies. Matching the features would required [C/Fe] = $+ 1.57$, compared to [C/Fe] = $+ 1.35$ as determined from the $G$-band. Importantly, increasing the total C abundance led to a significant overproduction of two $^{12}$CH lines at 4019.98\,{\AA} and 4020.13\,{\AA}, which suggests this scenario to not be correct. Similarly, we also increased the $^{13}$C/$^{12}$C ratio to match the observed spectrum rather than increasing the total C abundance. Again, we are unable to reproduce the two $^{12}$C lines around 4020\,{\AA} simultaneously with the $^{13}$C at the Th line position. All trial fits are shown in Figure~\ref{abundplot_Th}. We note that 
our $^{13}$C/$^{12}$C ratio, as derived from lines in the 4217\,{\AA} region, agrees with the ratio used for the fit of the Th 4019\,{\AA} region. Therefore, we conclude that  Th is indeed present in the star, and that our derived abundance is useful, within its stated uncertainties.

\startlongtable
\begin{deluxetable}{rrrr}
\tablecolumns{4}
\tablecaption{\label{synth_table} Line List of the $\lambda$4019 Region}
\tablehead{ 
\colhead{Species} & 
\colhead{$\lambda$ } & 
\colhead{$\chi$ } & 
\colhead{$\log gf$} }
\startdata
$^{13}$CH&4018.030&1.205&$-$4.554 \\
$^{13}$CH&4018.037&1.205&$-$2.737\\
$^{12}$CH&4018.045&1.001&$-$6.126\\
Ce\,II&4018.061&1.013&$-$0.500\\
Mn\,I&4018.063&3.378&$-$3.957\\
Mn\,I&4018.100&2.114&$-$0.309\\
V\,I&4018.110&6.334&$-$3.457\\
Th\,I&4018.121&0.000&$-$1.667\\
$^{13}$CH&4018.124&1.393&$-$6.256\\
Yb\,II&4018.134&5.341&$-$3.150\\
Ce\,II&4018.158&0.327&$-$2.270\\
$^{13}$CH&4018.158&1.393&$-$5.140\\
Co\,I&4018.160&3.576&$-$4.408\\
$^{13}$CH&4018.168&1.205&$-$2.660\\
$^{13}$CH&4018.175&1.205&$-$5.003\\
$^{12}$CH&4018.178&1.035&$-$3.926\\
Cr\,I&4018.205&2.708&$-$2.658\\
Yb\,II&4018.226&7.863&$-$2.750\\
$^{13}$CH&4018.243&1.393&$-$5.954\\
$^{12}$CH&4018.256&0.735&$-$9.130\\
Os\,I&4018.259&1.657&$-$1.230\\
Yb\,II&4018.262&7.251&$-$2.410\\
$^{12}$CH&4018.264&1.035&$-$4.782\\
Fe\,I&4018.267&3.266&$-$1.234\\
Yb\,II&4018.275&5.168&$-$3.110\\
Fe\,I&4018.325&3.686&$-$4.548\\
Mn\,I&4018.336&4.794&$-$2.444\\
$^{13}$CH&4018.347&0.732&$-$9.130\\
Zr\,II&4018.377&0.959&$-$1.270\\
Ce\,II&4018.386&0.875&$-$1.660\\
Ti\,II&4018.396&5.031&$-$2.202\\
Re\,I&4018.404&3.378&$-$0.270\\
$^{12}$CH&4018.419&1.035&$-$3.926\\
Fe\,II&4018.462&9.836&$-$3.786\\
Fe\,II&4018.490&2.276&$-$5.740\\
$^{12}$CH&4018.505&1.035&$-$4.793\\
Fe\,I&4018.506&4.209&$-$1.597\\
$^{12}$CH&4018.565&1.035&$-$3.828\\
Mn\,I&4018.570&5.087&$-$1.328\\
$^{12}$CH&4018.632&0.343&$-$3.842\\
Cu\,II&4018.686&14.423&$-$3.360\\
$^{12}$CH&4018.712&1.401&$-$6.256\\
Tm\,II&4018.737&3.349&$-$3.250\\
V\,I&4018.738&0.287&$-$6.805\\
$^{12}$CH&4018.766&1.401&$-$5.140\\
$^{12}$CH&4018.775&1.035&$-$3.828\\
F\,I&4018.800&12.985&$-$2.010\\
Ce\,II&4018.820&1.546&$-$0.960\\
Nd\,II&4018.823&0.064&$-$0.850\\
Cr\,I&4018.826&3.648&$-$2.629\\
$^{12}$CH&4018.835&1.401&$-$5.954\\
Cr\,I&4018.863&4.440&$-$2.822\\
$^{13}$CH&4018.888&1.206&$-$4.569\\
$^{13}$CH&4018.897&1.206&$-$2.737\\
Ce\,II&4018.900&1.013&$-$1.220\\
Ce\,II&4018.927&0.635&$-$1.680\\
V\,I&4018.929&2.581&$-$0.651\\
$^{12}$CH&4018.952&1.511&$-$2.349\\
Pr\,II&4018.963&0.204&$-$1.030\\
$^{13}$CH&4018.975&1.206&$-$2.660\\
$^{13}$CH&4018.976&0.462&$-$1.371\\
$^{13}$CH&4018.984&1.206&$-$4.989\\
U\,II&4018.986&0.036&$-$1.391\\
$^{12}$CH&4018.990&1.511&$-$4.755\\
Mn\,I&4018.999&4.354&$-$1.497\\
$^{13}$CH&4019.033&1.591&$-$2.147\\
$^{13}$CH&4019.035&0.462&$-$6.817\\
V\,I&4019.036&3.753&$-$2.704\\
Mn\,I&4019.042&4.666&$-$0.561\\
Fe\,I&4019.042&2.608&$-$2.780\\
$^{12}$CH&4019.052&1.511&$-$5.330\\
Ce\,II&4019.057&1.014&$-$0.390\\
Ni\,I&4019.058&1.935&$-$3.174\\
$^{12}$CH&4019.089&1.511&$-$2.437\\
$^{13}$CH&4019.103&1.591&$-$4.535\\
Fe\,II&4019.110&9.825&$-$3.102\\
Co\,I&4019.126&2.280&$-$2.270\\
$^{13}$CH&4019.126&1.591&$-$2.211\\
Th\,II&4019.129&0.000&$-$0.228\\
V\,I&4019.134&1.804&$-$1.999\\
Mo\,I&4019.143&3.399&$-$1.393\\
$^{13}$CH&4019.144&0.462&$-$1.336\\
Co\,I&4019.163&2.871&$-$3.136\\
Fe\,II&4019.181&7.653&$-$3.532\\
$^{13}$CH&4019.213&0.914&$-$3.775\\
W\,I&4019.228&0.412&$-$2.200\\
$^{12}$CH&4019.228&1.492&$-$4.444\\
Ce\,II&4019.271&0.328&$-$2.320\\
Cr\,II&4019.289&5.330&$-$5.604\\
Co\,I&4019.289&0.582&$-$3.232\\
Co\,I&4019.299&0.629&$-$3.769\\
Tm\,II&4019.309&4.174&$-$3.180\\
Yb\,II&4019.353&5.868&$-$0.720\\
$^{13}$CH&4019.356&1.503&$-$4.728\\
$^{13}$CH&4019.407&1.503&$-$2.348\\
Ni\,II&4019.422&6.329&$-$4.919\\
$^{12}$CH&4019.428&1.173&$-$8.930\\
V\,I&4019.448&2.583&$-$1.216\\
V\,I&4019.464&3.113&$-$2.893\\
Ce\,II&4019.471&0.875&$-$0.660\\
Cr\,I&4019.474&3.092&$-$5.391\\
Ca\,I&4019.483&2.933&$-$4.458\\
Ni\,II&4019.488&11.781&$-$3.093\\
V\,I&4019.490&2.582&$-$9.243\\
$^{13}$CH&4019.497&1.789&$-$8.046\\
Ce\,II&4019.498&1.079&$-$1.980\\
P\,I&4019.509&9.518&$-$1.908\\
Cr\,I&4019.514&4.402&$-$2.262\\
$^{13}$CH&4019.539&1.503&$-$2.437\\
Ru\,I&4019.543&1.063&$-$2.250\\
Mn\,I&4019.556&2.888&$-$3.141\\
$^{13}$CH&4019.590&1.503&$-$5.354\\
$^{12}$CH&4019.604&0.736&$-$9.175\\
Pb\,I&4019.632&2.660&$-$0.220\\
$^{13}$CH&4019.640&1.377&$-$9.731\\
C\,I&4019.645&7.946&$-$4.310\\
Ba\,II&4019.665&5.983&$-$3.510\\
$^{13}$CH&4019.674&0.733&$-$9.175\\
Mn\,II&4019.710&10.322&$-$2.042\\
Gd\,I&4019.726&0.066&$-$1.046\\
Cr\,I&4019.737&4.440&$-$1.379\\
$^{13}$CH&4019.741&1.789&$-$4.654\\
$^{13}$CH&4019.763&1.789&$-$4.582\\
Nd\,II&4019.810&0.631&$-$0.640\\
Sm\,II&4019.829&0.277&$-$1.530\\
Ni\,II&4019.879&13.125&$-$2.487\\
Tm\,II&4019.884&4.114&$-$1.560\\
Ce\,II&4019.897&1.014&$-$0.500\\
V\,I&4019.906&2.332&$-$4.856\\
Cr\,II&4019.941&11.483&$-$3.391\\
Sm\,II&4019.976&0.185&$-$1.250\\
$^{12}$CH&4019.982&1.209&$-$4.553\\
$^{12}$CH&4019.997&1.209&$-$2.737\\
$^{12}$CH&4020.019&0.465&$-$1.370\\
Ir\,I&4020.026&3.262&0.330\\
Fe\,I&4020.047&3.266&$-$4.590\\
Nd\,II&4020.051&1.272&$-$0.290\\
$^{13}$CH&4020.051&1.378&$-$6.094\\
Mn\,I&4020.068&3.771&$-$1.320\\
$^{12}$CH&4020.072&0.465&$-$6.817\\
$^{13}$CH&4020.088&1.377&$-$6.040\\
Mn\,I&4020.099&3.772&$-$7.702\\
$^{12}$CH&4020.125&1.209&$-$2.659\\
Yb\,II&4020.126&5.747&$-$3.760\\
$^{12}$CH&4020.140&1.209&$-$5.002\\
Fe\,I&4020.151&3.640&$-$4.070\\
Ce\,II&4020.154&1.027&$-$2.250\\
$^{12}$CH&4020.173&1.656&$-$8.520\\
$^{12}$CH&4020.186&0.465&$-$1.363\\
$^{13}$CH&4020.209&1.483&$-$4.444\\
Hf\,II&4020.250&1.780&$-$2.080\\
Ni\,I&4020.251&3.699&$-$0.936\\
Mn\,I&4020.283&4.268&$-$1.668\\
Yb\,II&4020.313&6.556&$-$2.150\\
Cr\,II&4020.329&11.249&$-$2.462\\
Sc\,I&4020.392&0.000&0.199\\
Cr\,I&4020.400&4.402&$-$3.723\\
Mo\,I&4020.402&3.739&$-$1.017\\
Cr\,II&4020.408&8.215&$-$4.019\\
Ti\,I&4020.428&2.239&$-$2.839\\
$^{12}$CH&4020.445&1.401&$-$4.511\\
Mo\,I&4020.451&2.758&$-$1.723\\
Tb\,II&4020.470&0.731&$-$0.130\\
Fe\,I&4020.484&3.642&$-$1.770\\
Er\,I&4020.513&0.863&0.595\\
Cr\,I&4020.519&4.204&$-$1.329\\
V\,I&4020.528&1.931&$-$3.099\\
Cr\,II&4020.528&11.249&$-$2.801\\
V\,I&4020.539&2.578&$-$2.655\\
Ce\,II&4020.541&0.635&$-$1.620
\enddata

\end{deluxetable}

Uncertainties of the neutron-capture element abundances range from 0.1 to 0.3\,dex,  depending on the level of blending and how well it can be accounted for. We take as uncertainties the standard error, as derived for small samples \citep{Keeping62}. In all cases, when the value is less than 0.1\,dex, we adopt 0.1\,dex as a more realistic minimum uncertainty. For elements with only one available line, we adopt an uncertainty between 0.1 and 0.3\,dex, depending on the quality of the measurement. Table~1 reports our final uncertainties for all elements.

Besides all the usual elements (e.g., Sr, Y, Zr, Ba, La, Ce, Nd, Sm, Eu, Gd, Dy, Pb) that can be measured in typical CEMP-$s$ stars, we were also able to measure Ru, Rh, and Pd, as well as Tm, Lu, Hf, Os, Ir and Th. Thorium is particularly noteworthy, as it immediately suggests that \thestar\ is not a 
purely $s$-process-enriched star, as Th is not synthesized in this process. 
Rather, the $r$-process makes thorium in large enough quantities that it can still be measured after $\sim$13\,Gyr. 
The $i$-process can also easily produce thorium. However, it is currently debated if the quantity would be sufficient for it to ever be observable as its production is likely several orders of magnitude below that of other neutron-capture elements (R. Standcliffe, priv. comm.). An $r$-process origin of the heaviest elements is furthermore suggested by the Os and Ir abundances. They are relatively high, matching the scaled $r$-process pattern (see Figure~4). The $i$-process is not expected to produce comparable amounts of third peak elements (M. Pignatari, priv. comm.).
The extremely high lead abundance stands in stark contrast. Pb is the end point of $s$-process nucleosynthesis and thus the heaviest element made this way. A large Pb abundance is the result $s$-process nucleosynthesis at low metallicity, where the available neutron-to-seed ratio is relatively large \citep{gallino1998,travaglio01}. This leads to an increased production of Pb, without similar amounts of e.g., third peak elements and lanthanides.

\section{The \textquotedblleft $r+s$" nature of \thestar} 

\subsection{Identification of the \textquotedblleft $r+s$" Pattern}
\begin{figure*}[!ht]
 \begin{center}
 \includegraphics[clip=true,width=14cm]{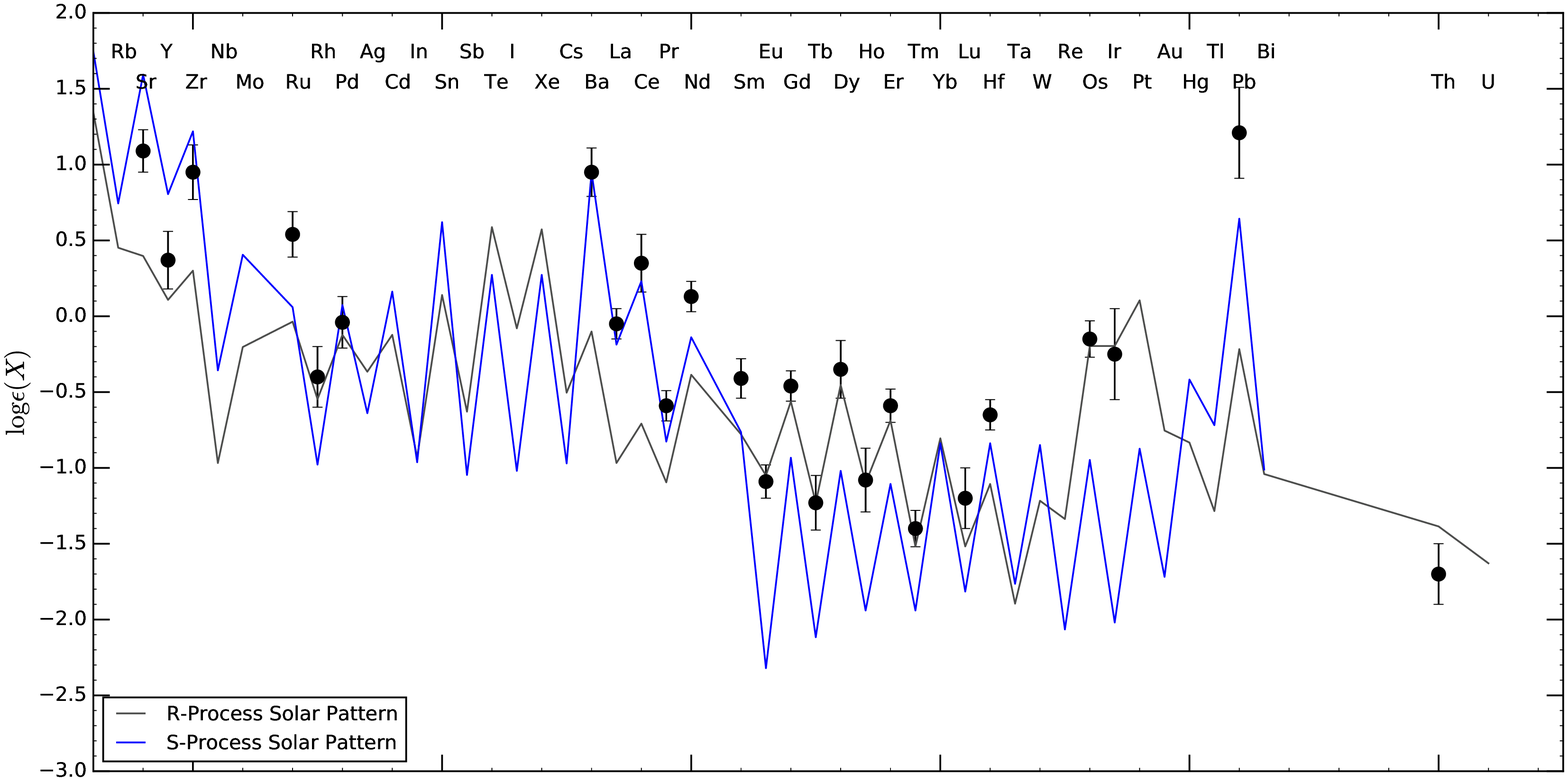} \\
 \includegraphics[clip=true,width=14cm]{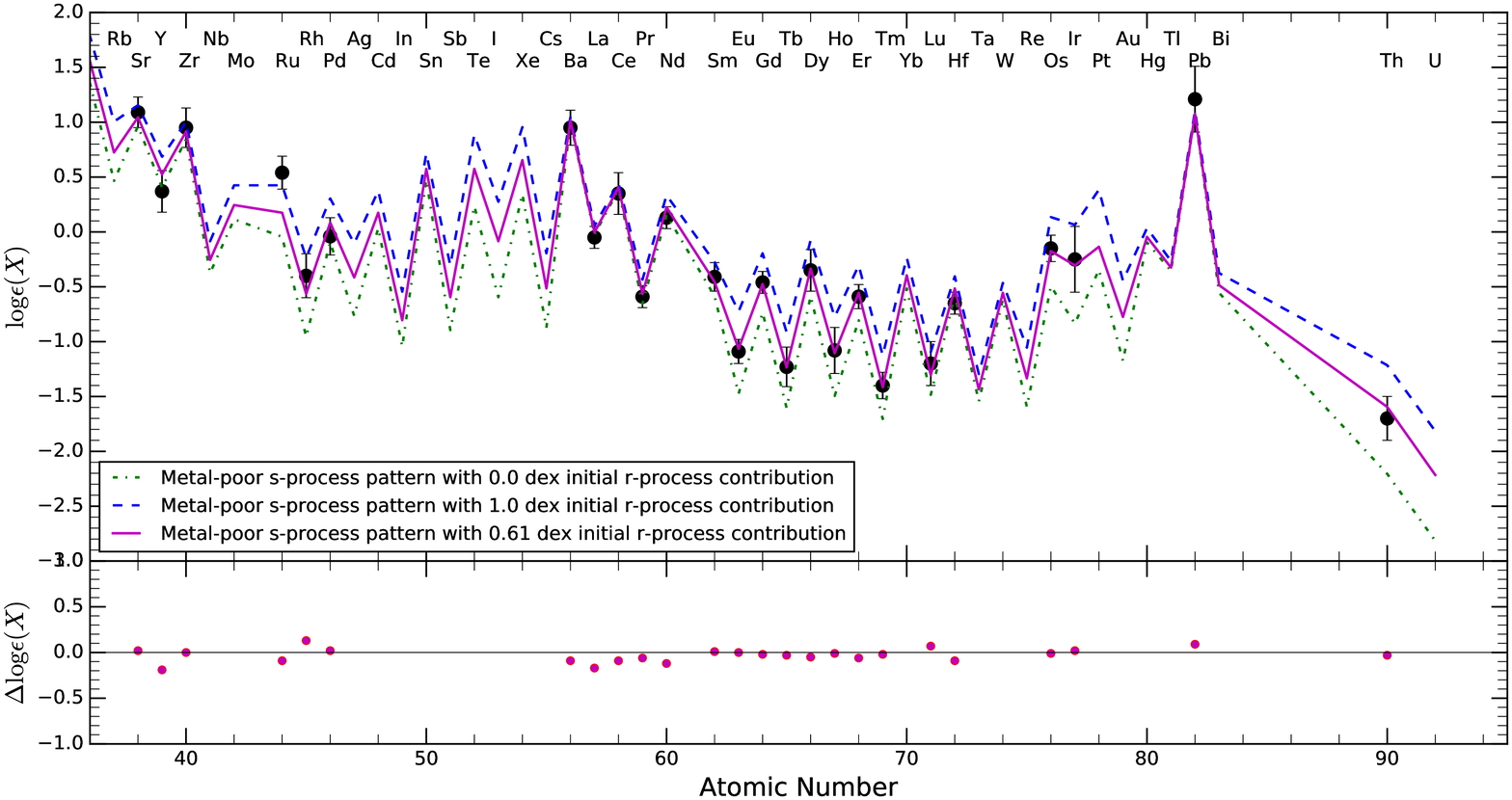}

    \figcaption{
     \label{abundplot} Neutron-capture element abundance ($\log \epsilon (\rm X)$), as a function of atomic number. Top panel: Abundances in comparison with the solar $s$-process pattern scaled to Ba and the solar $r$-process scaled to Eu. Neither pattern fits all the data. Bottom panel: Abundances in comparison with results from three metal-poor $s$-process models. The best fit (magenta line) is achieved with an $s$-process model combined with an initial $r$-process component of $\mbox{[Eu/Fe]} = + 0.6$. The other models have $r$-process contributions of $\mbox{[Eu/Fe]} =+ 0.0$ (blue line) and $\mbox{[Eu/Fe]} = + 1.0$ (green line).
Residuals, {\it i.e.} the difference between observations and the best-fit model, are shown in the bottom panel.}
 \end{center}
\end{figure*}

When comparing the neutron-capture abundances of \thestar\ with scaled solar $r$-process and $s$-process patterns, no clear match was found at first. Instead, light neutron-capture elements seemed to roughly follow the $s$-process pattern (with the $s$-process pattern scaled to Ba), while elements Eu and above instead followed an $r$-process pattern (when scaled to Eu). Figure~\ref{abundplot} shows this behavior in the top panel. The only exception was Pb. High Ba and Pb abundances clearly point to a low-metallicity $s$-process origin, but elements in between exhibit abundances that are too high for an $s$-process origin. In addition, the overall neutron-capture element pattern of \thestar\ does not resemble any of the CEMP-$i$ stars (e.g., \citealt{roederer2016}). Direct comparisons with models of \citet{Hampel16} also showed that the $i$-process is unlikely to have produced this pattern.
 
Since Th was detected in \thestar\ --  which indicates at least some contribution by an $r$-process to the natal gas cloud from which the star formed -- we resorted to combining, in a weighted fashion, the solar r- and $s$-process patterns. The weighting of the $r$-process pattern was based on the observed $\mbox{[Eu/Fe]}= + 0.6$ and that of the $s$-process was based on $\mbox{[Ba/Fe]}= + 0.99$. The result matches the overall neutron-capture element abundances fairly well. Inspired by the good fit, we then replaced the solar $s$-process pattern with a custom model of a 1.5\,M$_{\odot}$ star with $\mbox{[Fe/H]}=-2.3$ that produced $s$-process elements during its AGB phase \citep{Lugaro12}. This improved the fit again, now with the high Pb being precisely matched by the model.  Figure~\ref{abundplot} (bottom panel) shows this near-perfect fit of the abundances of \thestar\ made by this combination of $r$- and $s$-process patterns. For comparison, we also show models that have $r$-process contributions of $\mbox{[Eu/Fe]}=0.0$ and $\mbox{[Eu/Fe]}= + 1.0$. Differences are most apparent for the elements Os, Ir, and Th. These elements are particularly useful to constrain the $r$-process component, as the $s$-process contribution for these elements is comparatively little and none, respectively. 

All of the above confirms that the neutron-capture element abundances in \thestar\ are completely described by a combination of $s$-process and $r$-process nucleosynthesis. Also, assuming that the Eu abundance is largely due to the $r$-process, the overabundance of $\mbox{[Eu/Fe]}= + 0.6$ makes \thestar\ a moderately-enhanced $r$-I star, following the notation of \citet{ARAA}. Given that \thestar\ has, at face value, $\mbox{[Ba/Eu]}>0$, it would ordinarily not be classified as an $r$-process-enhanced star. This issue highlights that it is crucial to know about the entire abundance pattern in detail to ensure a physically meaningful interpretation of the observed chemical signature. 
Accordingly, given its high carbon abundance, it is likely that $r$-I star \thestar\ is in a binary star system with a companion that underwent its AGB phase and produced carbon and $s$-process elements. 
The natal gas from which the binary system must then have been enriched by an $r$-process event prior to its formation, likely by a neutron star merger \citep[e.g.,][]{Ji16b}.
While we cannot prove that the $r$-process components of the abundances in \thestar\ arose from gas enriched by a neutron star merger, no external pollution model can explain this chemical signature. Bondi-Hoyle accretion of $r$-process elements from the ISM affects stellar abundances typically at $10^{-6}$ of the solar metallicity level (e.g., \citealt{Komiya14, Shen17}), even neglecting the effect of stellar winds that would reduce the accretion rates.
Radial-velocity monitoring of halo $r$-process-enhanced stars suggests $r$-process pollution from a binary companion is unlikely to be important \citep{Hansen15}, and the only proposed mechanism for such pollution (a slow wind from the companion's electron-capture supernova, \citealt{wanajo_rs2006}) requires a massive companion that is inconsistent with our $s$-process models below.


\subsection{Modeling the Mass-Transfer of Carbon and $s$-Process Material}

Assuming that \thestar\ received its $s$-process component from a companion star, we also decided to model this putative mass-transfer event to gain insight into the binary system and the origin of its abundance pattern. 

In the binary mass-transfer scenario for the formation of CEMP stars, the primary star produces $s$-process elements in its interior, in the intershell region between the He- and the H-burning shells during its AGB phase. From this region, carbon and $s$-process elements are brought to the surface by recurrent deep convective episodes known as third dredge-ups. The products of internal nucleosynthesis dredged-up to the surface are subsequently released into the interstellar medium by a strong stellar wind. This material can then be partially accreted by the secondary, less-evolved star. To model all the relevant processes involved in this mass transfer scenario, we use the binary-evolution code \texttt{binary\_c/nucsyn} \citep{Izzard2004, Izzard2006, Izzard2009}. In particular, the chemical composition of the intershell region is computed as a function of three parameters: the mass of the star at the beginning of the AGB phase, the evolutionary stage along the AGB, and the mass of the partial mixing zone. The latter is a free parameter in the code, which determines the amount of free neutrons that are available in the intershell region for the production of $s$-process elements. A thorough description of the partial mixing zone and its role for $s$-process nucleosynthesis in AGB stars is provided by \citet{Karakas10}. \citet{Abate2015-1} describe the numerical treatment of the partial mixing zone used in \texttt{binary\_c/nucsyn}, and the method adopted to calculate the amount of material mixed to the surface by the third dredge-up, in order to reproduce the evolution predicted in the detailed AGB-nucleosynthesis models of \citet{Karakas10} and \citet{Lugaro12}. The efficiency of the accretion onto the secondary is then calculated according to the wind-Roche-lobe-overflow (WRLOF) model proposed by \citet[Eq. 9]{Abate2013} for a spherically symmetric wind. The transferred material is then diluted throughout the entire secondary star, by a combination of non-convective processes (such as diffusion and thermohaline mixing) and the first dredge-up, which occurs when the secondary star ascends the red giant branch.

We compare the observed abundances of \thestar\ with the grid of binary-star models computed by \citet{Abate2015-1}. The grid consists of about $285,\!000$ binary systems with initial parameters in a wide range of primary and secondary masses ($M_{1,\mathsf{i}}\in[0.9,6.0]M_{\odot}$, $M_{2,\mathsf{i}}\in[0.2,0.9]M_{\odot}$), orbital separations ($a_{\mathsf{i}}\in[10^2,10^5]R_{\odot}$), and masses of the partial mixing zone ($M_{\mathrm{PMZ}}\in[0.0,0.004]M_{\odot}$). The evolution of these binary systems is followed until both stars have become white dwarfs. We follow the method described by \citet{Abate2015-2} to determine the best-fit model to the observations. Initially, we constrain the evolutionary stage of the observed secondary by selecting from the grid of synthetic stars those that reproduce the measured
surface gravity within the observational uncertainty, at an
age $10 \leq t \leq 13.7$ Gyr (which is the likely age of halo stars). Subsequently, for the model stars that pass this selection, we determine how well they reproduce the observed abundances by computing the $\chi^2$ as follows:
\begin{equation}
\chi^2 = \sum_i\frac{(A_{i, \mathrm{obs}} - A_{i, \mathrm{mod}})^2}{\sigma^2_{i,\mathrm{obs}}}~,\label{eq:chi}
\end{equation}
where $A_{i, \mathrm{obs}}$ is the observed absolute abundance of element $i$, $A_{i, \mathrm{mod}}$ is the value predicted in the model, and $\sigma^2_{i,\mathrm{obs}}$ is the observational uncertainty. The minimum value of $\chi^2$ determines the best model. To calculate $\chi^2$ from Eq. (\ref{eq:chi}), we take into account all observed elements except those with atomic number between 14 (Si) and 29 (Cu). These elements are not produced by AGB nucleosynthesis, hence they are not useful to constrain the choice of our models as the differences with the observations arise from a discrepancy with our set of initial abundances (See Sections 4 and 5 of \citealt{Abate2015-2}).

\begin{figure*}[!ht]
 \begin{center}
  \includegraphics[clip=true,width=16cm]{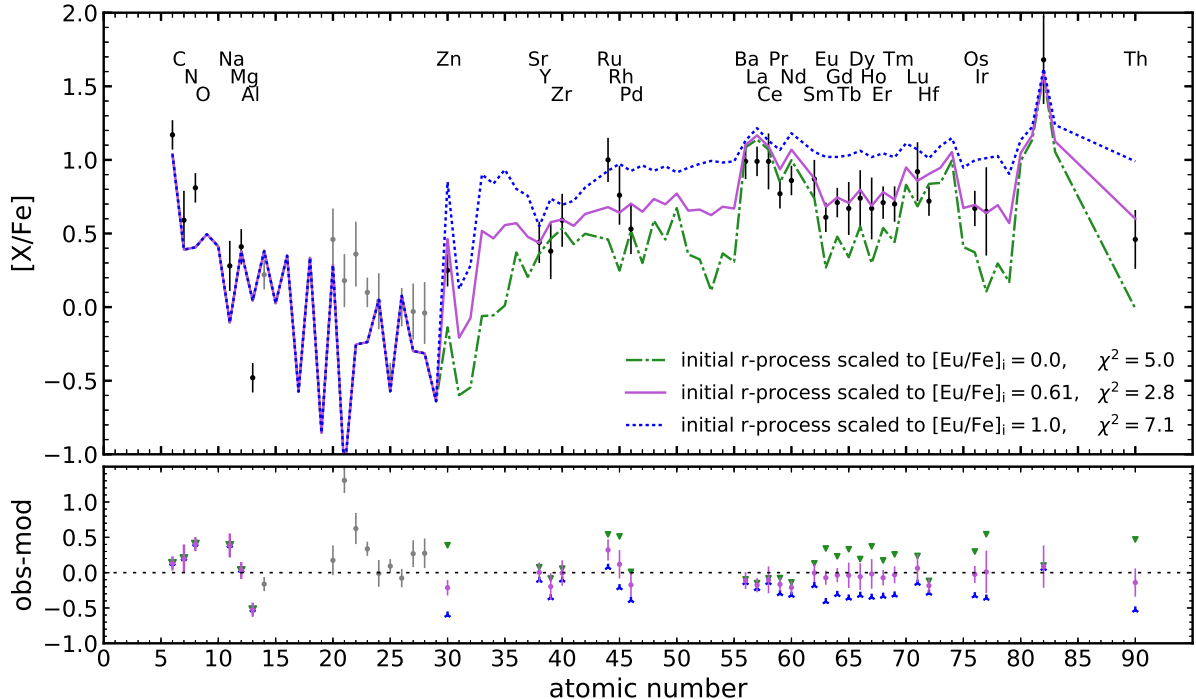} 
    \figcaption{Complete abundance pattern of \thestar\, as a function of atomic number, overlaid with predictions by several binary-evolution and nucleosynthesis models of the surface composition of the star after the mass transfer. The $s$-process model with an initial $r$-process enhancement scaled to $\mbox{[Eu/Fe]}= + 0.6$ fits the data best. Residuals are shown at the bottom. 
     \label{carlo}}
 \end{center}
\end{figure*}

In our best-fit model the initial binary system consisted of a $0.9M_{\odot}$ primary and a $0.86M_{\odot}$ secondary star in a $4634$-day orbit. The mass of the partial mixing zone during the AGB phase of the primary star is $M_{\mathrm{PMZ}}=0.001M_{\odot}$. According to the model, the secondary star accreted $0.1M_{\odot}$ of material when the donor was in its AGB phase. The  period of the current binary system (in which the secondary star is a now a carbon-enhanced red giant, while the erstwhile primary is an unseen white dwarf) is approximately 5590 days or 15.3 years. The results are shown in Figure~\ref{carlo}. This qualitatively agrees with the non-detection of radial-velocity variations for \thestar\ to within a few km\,s$^{-1}$.

We also varied the input to the fit, such as the number of elements used and the initial $r$-process abundance level (see more below). The binary parameters did not significantly change during these tests, suggesting that our overall results are relatively robust. Increasing the mass of the primary star and of the partial mixing zone during the AGB phase causes a significant increase in the abundances of light elements such as C, Na, Mg, and of heavy $s$-process elements from Ba to Pb. However, this is at odds with the observed abundances, and consequently causes the $\chi^2$ of the fit to increase. Hence, we regard our mass estimate and assumed partial mixing zone resulting from our best fit as our final, robust values.

In our default model (green dot-dashed line in Figure~\ref{carlo}), we neglect that other neutron-capture element sources may have enriched the gas cloud from which \thestar\ formed. However, as  discussed above, the [Th/Fe] ratio is three times as high as in the Sun, suggesting that the binary system formed from gas that had previously been enriched in $r$-process elements. To confirm this hypothesis, we compute two additional models in which we assume such an initial $r$-process enhancement. This $r$-process component is calculated by scaling the abundances of all neutron-capture elements from Zn to U to the observed Eu abundance, $[\mathrm{Eu/Fe}]= + 0.6$ (assuming an $r$-process pattern), and also a second test case with $[\mathrm{Eu/Fe}]= + 1.0$. 

These two models are shown in Figure~\ref{carlo} as solid and dotted lines, respectively. Elements in the first $s$-process peak (Sr, Y, Zr), in the second $s$-process peak (Ba, La, Ce), and also lead (third peak), are abundantly produced during AGB nucleosynthesis, therefore their final enhancement is hardly affected by their initial abundances. In contrast, the abundances of elements typically associated with the $r$-process (most of the elements heavier than Nd and up to Pb) are produced only in small amounts during AGB nucleosynthesis. Consequently, the addition of our initial $r$-process component changes the final abundances. The observed abundances of these elements are much better reproduced by our model with initial abundances scaled to $[\mathrm{Eu/Fe}]=+ 0.6$. The model with the initial $[\mathrm{Eu/Fe}]=+ 1.0$ over-produced the observed abundances of \thestar,  showing that the initial $r$-process enrichment can be very well-constrained by the observations. This confirms \thestar\ as an $r$-I star.

\section{The age of \thestar}

Given that \thestar\ is an r-I star and thorium was detected,
we attempted to measure its age through cosmo-chronometry. Assuming that the contribution to Eu by the $s$-process is negligible, we choose the Th/Eu ratio for the age determination. We also considered Os and Ir, as the $s$-process contributions are minor, especially in the case of Ir. However, Os and especially Ir measurements generally have larger uncertainties than Eu, which is of significance when using them for age determinations. Other neutron-capture elements are more contaminated by the $s$-process contribution to the overall abundance pattern, thus we refrain from using them for the age dating.

We employ 
$\Delta t = 46.78 [\log (\rm{Th/r})_{\rm initial} - \log \epsilon(Th/r)_{now}]$ \citep{Cayreletal:2001} to derive the age of \thestar. The $\log (\rm{Th/r})_{\rm initial}$ refers to the ratio of Th to a stable $r$-process element produced in the original nucleosynthesis event.
In terms of initial production ratios, we use values from \citet{schatz02}, 
$\log (\rm{Th/Eu})_{\rm initial} = -0.33$, 
$\log (\rm{Th/Os})_{\rm initial} = -1.15$, 
$\log (\rm{Th/Ir})_{\rm initial} = -1.18$. 
Taking abundances from Table~1, we obtain 
$\log \epsilon(\rm{Th/Eu}) = -0.61$, 
$\log \epsilon(\rm{Th/Os}) = -1.55$, 
$\log \epsilon(\rm{Th/Ir}) = -1.45$. 
This translates into ages of 
13.1\,Gyr from Th/Eu, 
18.7\,Gyr from Th/Os, and 
12.6\,Gyr from Th/Ir. 
We adopt the Th/Eu-based age of 13.1\,Gyr as our final age estimate of \thestar. We note that adopting the WP1 model initial production ratios from \citet{hill17} yields ages of 11.0, 19.9, and 13.9\,Gyr. We note here that the Th/Eu ratio can be affected by an actinide-boost. However, given that 75\% of r-process stars are not actinide-boost stars \citep{mashonkina14}, we can fairly assume that \thestar is not affected. In fact, all actinide boost stars yield significantly young or even negative ages, making them relatively easily identifiable by the age their Th/Eu ratio supplies. The age of J0949-1617 of 13.1 Gyr does not fall into this category. It neither suggests a result similar to that of the brightest star in Reticulum\,II \citep{ji18} which appears to be actinide-deficient and thus yields a supposed age of 22\,Gyr or older.

Age uncertainties are generally large. Measurement uncertainties of 0.05\,dex translate to an uncertainty of 2.3\,Gyr for Th/$r$-elements ratios. We consider this an optimistic uncertainty -- likely it  is of order 5\,Gyr. Given that this is already rather large, we do not pursue additional sources of error. Taking this into account, the ages derived from the Th/Os and Th/Ir ratios agree with the value obtained from the Th/Eu ratio. Overall, despite the uncertainties, these values confirm that \thestar\ is an old star.


\section{Conclusion}

We have discovered that the metal-poor giant \thestar\ is the first true  \textquotedblleft $r+s$" star, i.e. a CEMP-$s$ star that formed from $r$-process enriched gas. In fact, \thestar\ is a moderately-enhanced $r$-I star, with $\mbox{[Eu/Fe]}\sim + 0.6$. 

Assuming that \thestar\ is a member of a binary companion, we modeled the mass transfer of $s$-process elements from a former companion that went through the AGB phase onto the presently observed star. The binary mass-transfer scenario is supported by the large carbon over-abundance of $\mbox{[C/Fe]}\sim + 1.2$. 
With our binary evolution and nucleosynthesis code we find that the best fit to the observed abundances has primary and secondary masses equal to $M_{1,\mathrm{i}} = 0.9M_{\odot}$ and $M_{2,\mathrm{i}} =0.86M_{\odot}$, respectively, mass of the partial mixing zone $M_{\mathrm{PMZ}}=0.001M_{\odot}$, and initial and final orbital periods equal to $P_{\mathrm{i}} = 4634$ and $P_{\mathrm{i}} = 5590$ days, respectively.

We note for completeness that stars with abundance signatures seemingly arising from a combination of the $r$ and $s$-process might also be explained with a single site, namely 20-30\,M$_{\odot}$ low-metallicity stars \citep{banerjee17}. They might host a neutron-capture site that would lead to $s$- and $i$-process nucleosynthesis. It remains to be seen whether \thestar\ could be explainable in this way, as this process could only produce elements up to Bi ($Z =83$), leaving the origin of Th ($Z=90$) unaccounted for.

Commensurate with expectations for $r$-process-enhanced stars, Th was detected in the spectrum of \thestar. Its abundance is consistent with that of other $r$-process-enhanced stars at similar metallicities (e.g., \citealt{Ren2012}). This is independent evidence that a (main) $r$-process must have enriched the gas before the star formed. The main $r$-process, which produces heavy neutron-capture elements from Ba up to Th and U, is now believed to occur primarily in neutron star mergers (or possibly magneto-rotationally driven jet supernovae) but not ordinary core-collapse supernovae \citep{wanajo01}. Given recent results by \citet{Ji16b}, who suggested that a neutron star merger enriched the ultra-faint dwarf galaxy Reticulum\,II, we speculate that \thestar\ must have formed in an environment that was also enriched by a neutron star merger. Given that the level of $r$-process enhancement is about 1\,dex lower than that found in Ret\,II, the birth system of \thestar\ would have likely been of order 10 times more massive, so that any $r$-process yield would have been sufficiently diluted prior to the formation of \thestar. Such an environment could resemble that of the ultra-faint dwarf Tucana\,III, which contains $r$-I stars \citep{hansen17} at a similar enhancement level as \thestar. 

Future searches for metal-poor stars will hopefully soon uncover more of these \textquotedblleft $r+s$" stars. Assuming a 15\% occurrence rate of both $r$-I and $s$-process-enhanced stars would suggest an expected rate of $r+s$ stars of 2-3\%. \citet{Barklem05} found a rate of 3\% for the frequency of $r$-II stars, and about two dozen $r$-II stars are known today, despite their rarity. Is it thus clear that $r+s$ stars must be more rare than $r$-II stars. This somewhat surprising paucity strongly suggests a low occurrence rate of binary stars in the earliest galaxies that were enriched by rare neutron star merger events.


\startlongtable
\begin{deluxetable}{lrrrrrrrrrrrrrrrrrrrr}
\tablecolumns{6}
\tabletypesize{\footnotesize}
\tabletypesize{\tiny}
\tablewidth{0pc}
\tablecaption{\label{Equivalent Widths} Equivalent width measurements of \thestar}
\tablehead{ \colhead{Element} &
\colhead{$\lambda$} & \colhead{EP} & \colhead{$\log$ \textit{gf}}& \colhead{EW}  & \colhead{$\log\epsilon$(X)} \\
\colhead{}& \colhead{[\AA]} & \colhead{[eV]} & \colhead{dex} & \colhead{[m\AA]} & \colhead{[dex]}}
\startdata
C (CH)& 4312     & ...  & ...     & syn   & 7.37\\
C (CH)& 4323     & ...  & ...     & syn   & 7.38\\
N (NH)& 3360     & ...  & ...     & syn   & 6.20 \\
O\,I  & 6300.30  & 0.00 & $-$9.82 & 8.86  & 7.25\\
O\,I  & 6363.80  & 0.02 & $-$10.30& 3.31  & 7.31\\
Na\,I & 5682.60  & 2.10 & $-$0.70 & 9.73  & 4.25\\
Na\,I & 5688.20  & 2.10 & $-$0.45 & 10.79 & 4.05 \\
Na\,I & 5890.00  & 0.00 & 0.11    & 211.85& 4.43\\
Na\,I & 5895.90  & 0.00 & $-$0.19 & 189.83& 4.47\\
Mg\,I & 3829.40  & 2.71 & $-$0.23 & 283.14& 5.97\\
Mg\,I & 3832.30  & 2.71 & 0.27    & 361.49& 5.75\\
Mg\,I & 3986.80  & 4.35 & $-$1.06 & 75.67 & 5.99\\
Mg\,I & 4571.10  & 0.00 & $-$5.69 & 90.03 & 5.81\\
Mg\,I & 4703.00  & 4.33 & $-$0.38 & 100.77& 5.70\\
Mg\,I & 5172.70  & 2.71 & $-$0.39 & 231.15& 5.67\\
Mg\,I & 5183.60  & 2.72 & $-$0.16 & 267.82& 5.68\\
Mg\,I & 5528.40  & 4.35 & $-$0.50 & 102.33& 5.81\\
Mg\,I & 5711.10  & 4.34 & $-$1.72 & 24.49 & 5.75\\
Al\,I & 3961.50  & 0.01 & $-$0.33 & 157.46& 3.75\\
Si\,I & 5665.60  & 4.90 & $-$1.73 & 5.54  & 5.59\\
Si\,I & 5701.10  & 4.93 & $-$2.05 & 2.30  & 5.54\\
Si\,I & 3906.52  & 1.91 & $-$1.09 & syn   & 5.40 \\
Ca\,I & 4318.70  & 1.90 & $-$0.21 & 82.50 & 4.39\\
Ca\,I & 4425.40  & 1.88 & $-$0.36 & 100.64& 4.89 \\
Ca\,I & 4435.70  & 1.88 & $-$0.52 & 103.60& 5.12 \\
Ca\,I & 4454.78  & 1.90 & 0.25    & 129.01& 4.87\\
Ca\,I & 4455.89  & 1.90 & $-$0.52 & 73.72 & 4.51\\
Ca\,I & 5262.24  & 2.52 & $-$0.47 & 64.68 & 4.95\\
Ca\,I & 5265.56  & 2.52 & $-$0.26 & 72.68 & 4.88\\
Ca\,I & 5349.46  & 2.71 & $-$0.31 & 27.30 & 4.34\\
Ca\,I & 5581.96  & 2.52 & $-$0.56 & 32.48 & 4.47\\
Ca\,I & 5588.75  & 2.52 & 0.21    & 76.22 & 4.46\\
Ca\,I & 5590.11  & 2.52 & $-$0.57 & 27.13 & 4.38\\
Ca\,I & 5594.46  & 2.52 & 0.10    & 67.27 & 4.42\\
Ca\,I & 5598.48  & 2.52 & $-$0.09 & 63.54 & 4.53\\
Ca\,I & 5601.28  & 2.53 & $-$0.52 & 37.03 & 4.53 \\
Ca\,I & 5857.45  & 2.93 & 0.23    & 46.85 & 4.40\\
Ca\,I & 6102.72  & 1.88 & $-$0.79 & 60.57 & 4.43\\
Ca\,I & 6122.22  & 1.88 & $-$0.32 & 93.76 & 4.53\\
Ca\,I & 6162.17  & 1.90 & $-$0.09 & 109.66& 4.61\\
Ca\,I & 6169.04  & 2.52 & $-$0.80 & 21.75 & 4.46\\
Ca\,I & 6169.56  & 2.53 & $-$0.48 & 34.43 & 4.42\\
Ca\,I & 6439.08  & 2.52 & 0.47    & 90.40 & 4.40\\
Ca\,I & 6449.81  & 2.52 & $-$0.50 & 45.07 & 4.61\\
Ca\,I & 6717.68  & 2.71 & $-$0.52 & 33.86 & 4.64\\
Sc\,II & 4246.82 & 0.60 & 0.24    & syn   & 1.00 \\
Sc\,II & 4314.08 & 0.62 & $-$0.10 & syn   & 1.00 \\
Sc\,II & 4325.00 & 0.59 & $-$0.44 & syn   & 1.08 \\
Sc\,II & 4400.39 & 0.61 & $-$0.54 & syn   & 1.14 \\
Sc\,II & 4415.54 & 0.60 & $-$0.67 & syn   & 1.09 \\
Sc\,II & 5031.01 & 1.36 & $-$0.40 & syn   & 0.92 \\
Sc\,II & 5318.37 & 1.36 & $-$2.01 & syn   & 1.09 \\
Sc\,II & 5526.78 & 1.77 & 0.02    & syn   & 0.92 \\
Sc\,II & 5641.00 & 1.50 & $-$1.13 & syn   & 1.50 \\
Sc\,II & 5658.36 & 1.49 & $-$1.21 & syn   & 1.23 \\
Sc\,II & 5667.16 & 1.50 & $-$1.31 & syn   & 1.25 \\
Sc\,II & 5669.06 & 1.50 & $-$1.20 & syn   & 1.00 \\
Sc\,II & 5684.21 & 1.51 & $-$1.07 & syn   & 0.90 \\
Sc\,II & 6604.60 & 1.36 & $-$1.31 & syn   & 1.48 \\
Ti\,I & 3904.78  & 0.90 & 0.15    & 41.96 & 2.76\\
Ti\,I & 3989.76  & 0.02 & $-$0.13 & 98.14 & 3.17 \\
Ti\,I & 3998.64  & 0.05 & 0.02    & 84.26 & 2.70\\
Ti\,I & 4008.93  & 0.02 & $-$1.00 & 48.21 & 2.98\\
Ti\,I & 4512.73  & 0.84 & $-$0.40 & 38.17 & 3.09 \\
Ti\,I & 4518.02  & 0.83 & $-$0.25 & 44.45 & 3.04 \\
Ti\,I & 4533.24  & 0.85 & 0.54    & 73.83 & 2.79\\
Ti\,I & 4534.78  & 0.84 & 0.35    & 65.34 & 2.81\\
Ti\,I & 4535.57  & 0.83 & 0.14    & 63.01 & 2.97\\
Ti\,I & 4548.76  & 0.83 & $-$0.28 & 35.70 & 2.91\\
Ti\,I & 4555.48  & 0.85 & $-$0.40 & 26.57 & 2.88\\
Ti\,I & 4656.47  & 0.00 & $-$1.35 & 35.46 & 2.99\\
Ti\,I & 4681.91  & 0.05 & $-$1.07 & 54.16 & 3.09 \\
Ti\,I & 4840.87  & 0.90 & $-$0.43 & 30.28 & 3.02 \\
Ti\,I & 4981.73  & 0.85 & 0.57    & 81.50 & 2.85\\
Ti\,I & 4991.07  & 0.84 & 0.45    & 80.66 & 2.93\\
Ti\,I & 4999.50  & 0.83 & 0.32    & 82.58 & 3.09 \\
Ti\,I & 5007.21  & 0.82 & 0.17    & 78.48 & 3.15 \\
Ti\,I & 5014.19  & 0.00 & $-$1.22 & 82.34 & 3.64\\
Ti\,I & 5014.28  & 0.81 & 0.04    & 84.24 & 3.38\\
Ti\,I & 5016.16  & 0.85 & $-$0.48 & 28.01 & 2.95\\
Ti\,I & 5020.03  & 0.84 & $-$0.33 & 31.56 & 2.86\\
Ti\,I & 5024.84  & 0.82 & $-$0.53 & 25.20 & 2.91\\
Ti\,I & 5035.90  & 1.46 & 0.22    & 29.77 & 3.00 \\
Ti\,I & 5036.46  & 1.44 & 0.14    & 22.97 & 2.91\\
Ti\,I & 5038.40  & 1.43 & 0.02    & 20.13 & 2.95\\
Ti\,I & 5039.96  & 0.02 & $-$1.08 & 45.69 & 2.90\\
Ti\,I & 5064.65  & 0.05 & $-$0.94 & 57.49 & 2.98\\
Ti\,I & 5173.74  & 0.00 & $-$1.06 & 54.09 & 2.98\\
Ti\,I & 5192.97  & 0.02 & $-$0.95 & 60.90 & 2.99\\
Ti\,I & 5210.38  & 0.05 & $-$0.82 & 61.52 & 2.91\\
Ti\,I & 6258.10  & 1.44 & $-$0.39 & 11.37 & 3.01 \\
Ti\,II & 3383.76 & 0.00 & 0.15    & 334.22& 2.93\\
Ti\,II & 3387.83 & 0.03 & $-$0.41 & 216.41& 3.02 \\
Ti\,II & 3394.57 & 0.01 & $-$0.54 & 214.01& 3.12 \\
Ti\,II & 3456.38 & 2.06 & $-$0.11 & 81.65 & 2.82\\
Ti\,II & 3477.18 & 0.12 & $-$0.95 & 195.31& 3.50\\
Ti\,II & 3491.05 & 0.11 & $-$1.10 & 142.16& 3.02 \\
Ti\,II & 3759.29 & 0.61 & 0.28    & 223.24& 2.88\\
Ti\,II & 3761.32 & 0.57 & 0.18    & 208.51& 2.84\\
Ti\,II & 3913.46 & 1.12 & $-$0.36 & 150.80& 3.29\\
Ti\,II & 4012.38 & 0.57 & $-$1.78 & 110.85& 3.15 \\
Ti\,II & 4025.13 & 0.61 & $-$2.11 & 113.99& 3.60\\
Ti\,II & 4161.53 & 1.08 & $-$2.09 & 89.91 & 3.50\\
Ti\,II & 4163.64 & 2.59 & $-$0.13 & 72.28 & 2.91\\
Ti\,II & 4395.84 & 1.24 & $-$1.93 & 78.24 & 3.23\\
Ti\,II & 4417.72 & 1.16 & $-$1.43 & 124.44& 3.66\\
Ti\,II & 4418.33 & 1.24 & $-$1.99 & 83.87 & 3.39\\
Ti\,II & 4441.73 & 1.17 & $-$2.41 & 66.82 & 3.41\\
Ti\,II & 4443.80 & 1.08 & $-$0.71 & 131.51& 2.99 \\
Ti\,II & 4444.55 & 1.12 & $-$2.20 & 76.95 & 3.31\\
Ti\,II & 4450.48 & 1.08 & $-$1.52 & 106.71& 3.24\\
Ti\,II & 4464.45 & 1.16 & $-$2.08 & 91.08 & 3.53\\
Ti\,II & 4468.49 & 1.13 & $-$0.63 & 147.86& 3.30\\
Ti\,II & 4470.86 & 1.16 & $-$2.28 & 73.39 & 3.38\\
Ti\,II & 4488.32 & 3.12 & $-$0.50 & 27.22 & 3.01 \\
Ti\,II & 4493.52 & 1.08 & $-$2.78 & 34.93 & 3.10 \\
Ti\,II & 4501.27 & 1.12 & $-$0.77 & 129.21& 3.02 \\
Ti\,II & 4529.48 & 1.57 & $-$1.75 & 63.75 & 3.14 \\
Ti\,II & 4533.97 & 1.24 & $-$0.77 & 157.38& 3.71\\
Ti\,II & 4545.13 & 1.13 & $-$2.45 & 52.52 & 3.13 \\
Ti\,II & 4563.76 & 1.22 & $-$0.96 & 124.38& 3.20\\
Ti\,II & 4571.97 & 1.57 & $-$0.31 & 140.53& 3.31\\
Ti\,II & 4583.41 & 1.16 & $-$2.84 & 27.29 & 3.11 \\
Ti\,II & 4589.96 & 1.24 & $-$1.79 & 85.15 & 3.18\\
Ti\,II & 4657.20 & 1.24 & $-$2.29 & 47.67 & 3.01 \\
Ti\,II & 4708.66 & 1.24 & $-$2.35 & 54.16 & 3.16 \\
Ti\,II & 4779.98 & 2.05 & $-$1.37 & 54.64 & 3.13 \\
Ti\,II & 4798.53 & 1.08 & $-$2.66 & 51.48 & 3.24\\
Ti\,II & 4805.08 & 2.06 & $-$1.10 & 73.69 & 3.21 \\
Ti\,II & 4865.61 & 1.12 & $-$2.70 & 38.09 & 3.09 \\
Ti\,II & 5005.17 & 1.57 & $-$2.73 & 23.03 & 3.33\\
Ti\,II & 5129.16 & 1.89 & $-$1.34 & 85.02 & 3.43\\
Ti\,II & 5185.90 & 1.89 & $-$1.41 & 57.99 & 3.015 \\
Ti\,II & 5188.68 & 1.58 & $-$1.21 & 122.62& 3.68\\
Ti\,II & 5226.54 & 1.57 & $-$1.30 & 92.26 & 3.12 \\
Ti\,II & 5268.61 & 2.60 & $-$1.61 & 11.40 & 3.02 \\
Ti\,II & 5336.79 & 1.58 & $-$1.60 & 73.04 & 3.08 \\
Ti\,II & 5381.02 & 1.57 & $-$1.97 & 53.19 & 3.10 \\
Ti\,II & 5418.77 & 1.58 & $-$2.13 & 44.97 & 3.14 \\
V\,II & 3951.41  & 1.48 & $-$0.78 & syn   & 1.70 \\
V\,II & 4005.71  & 1.82 & $-$0.52 & syn   & 1.88 \\
V\,II & 4023.38  & 1.80 & $-$0.61 & 43.99 & 1.88 \\
V\,II & 4380.54  & 2.14 & $-$0.23 & syn   & 1.77 \\
Cr\,I & 3578.70  & 0.00 & 0.42    & 152.42& 3.27\\
Cr\,I & 3908.76  & 1.00 & $-$1.05 & 34.45 & 3.13 \\
Cr\,I & 4254.35  & 0.00 & $-$0.09 & 143.69& 3.27\\
Cr\,I & 4545.95  & 0.94 & $-$1.37 & 32.68 & 3.27\\
Cr\,I & 4580.05  & 0.94 & $-$1.66 & 34.35 & 3.59\\
Cr\,I & 4600.75  & 1.00 & $-$1.25 & 35.93 & 3.28 \\
Cr\,I & 4616.12  & 0.98 & $-$1.19 & 40.62 & 3.28\\
Cr\,I & 4626.17  & 0.97 & $-$1.33 & 29.63 & 3.19\\
Cr\,I & 4646.16  & 1.03 & $-$0.74 & 56.01 & 3.14 \\
Cr\,I & 4651.29  & 0.98 & $-$1.46 & 27.77 & 3.30\\
Cr\,I & 4652.16  & 1.00 & $-$1.04 & 47.98 & 3.28\\
Cr\,I & 5206.02  & 0.94 & 0.02    & 109.42& 3.26 \\
Cr\,I & 5247.56  & 0.96 & $-$1.59 & 23.56 & 3.27\\
Cr\,I & 5296.69  & 0.98 & $-$1.36 & 32.45 & 3.25\\
Cr\,I & 5298.27  & 0.98 & $-$1.14 & 43.62 & 3.23\\
Cr\,I & 5300.75  & 0.98 & $-$2.00 & 10.51 & 3.29\\
Cr\,I & 5345.80  & 1.00 & $-$0.95 & 56.34 & 3.27\\
Cr\,I & 5348.31  & 1.00 & $-$1.21 & 35.38 & 3.17 \\
Cr\,I & 5409.78  & 1.03 & $-$0.67 & 67.29 & 3.19\\
Cr\,II & 4558.65 & 4.07 & $-$0.66 & 46.07 & 3.71\\
Cr\,II & 4588.20 & 4.07 & $-$0.83 & 32.54 & 3.62 \\
Mn\,I & 4041.36 & 2.11 & 0.29    & syn   & 2.69 \\
Mn\,I & 4754.05 & 2.28 & $-$0.09 & syn   & 2.75 \\
Mn\,I & 4783.43 & 2.30 & 0.04    & syn   & 2.76 \\
Fe\,I & 4430.61  & 2.22 & $-$1.73 & 76.63 & 5.42\\
Fe\,I & 4443.19  & 2.86 & $-$1.02 & 58.82 & 5.11 \\
Fe\,I & 4454.38  & 2.83 & $-$1.25 & 56.55 & 5.26\\
Fe\,I & 4461.65  & 0.09 & $-$3.19 & 127.17& 5.50\\
Fe\,I & 4466.55  & 2.83 & $-$0.59 & 90.50 & 5.29\\
Fe\,I & 4484.22  & 3.60 & $-$0.64 & 32.45 & 5.10 \\
Fe\,I & 4489.74  & 0.12 & $-$3.90 & 92.67 & 5.43\\
Fe\,I & 4490.08  & 3.02 & $-$1.58 & 23.64 & 5.18 \\
Fe\,I & 4494.56  & 2.19 & $-$1.14 & 99.80 & 5.31\\
Fe\,I & 4531.15  & 1.48 & $-$2.10 & 93.33 & 5.27\\
Fe\,I & 4592.65  & 1.56 & $-$2.46 & 81.03 & 5.43\\
Fe\,I & 4602.00  & 1.61 & $-$3.13 & 37.53 & 5.37\\
Fe\,I & 4602.94  & 1.48 & $-$2.21 & 93.49 & 5.37\\
Fe\,I & 4607.65  & 3.26 & $-$1.33 & 26.09 & 5.26\\
Fe\,I & 4619.29  & 3.60 & $-$1.06 & 26.36 & 5.38\\
Fe\,I & 4625.04  & 3.24 & $-$1.27 & 33.69 & 5.33\\
Fe\,I & 4630.12  & 2.28 & $-$2.59 & 27.88 & 5.42\\
Fe\,I & 4632.91  & 1.61 & $-$2.90 & 45.59 & 5.28\\
Fe\,I & 4637.50  & 3.28 & $-$1.29 & 21.50 & 5.13 \\
Fe\,I & 4643.46  & 3.64 & $-$1.15 & 14.87 & 5.20\\
Fe\,I & 4647.43  & 2.95 & $-$1.31 & 43.06 & 5.20\\
Fe\,I & 4668.13  & 3.26 & $-$1.08 & 40.28 & 5.29\\
Fe\,I & 4669.17  & 3.65 & $-$1.25 & 14.27 & 5.29\\
Fe\,I & 4678.85  & 3.60 & $-$0.68 & 40.40 & 5.28\\
Fe\,I & 4691.41  & 2.99 & $-$1.45 & 39.20 & 5.32\\
Fe\,I & 4707.27  & 3.24 & $-$0.96 & 50.14 & 5.32\\
Fe\,I & 4710.28  & 3.02 & $-$1.61 & 36.72 & 5.47\\
Fe\,I & 4733.59  & 1.48 & $-$2.99 & 55.25 & 5.38\\
Fe\,I & 4772.80  & 1.56 & $-$2.90 & 37.09 & 5.06 \\
Fe\,I & 4786.81  & 3.00 & $-$1.61 & 34.76 & 5.40 \\
Fe\,I & 4789.65  & 3.53 & $-$0.96 & 30.37 & 5.27\\
Fe\,I & 4859.74  & 2.87 & $-$0.85 & 86.21 & 5.44\\
Fe\,I & 4871.32  & 2.87 & $-$0.34 & 99.99 & 5.22\\
Fe\,I & 4872.14  & 2.88 & $-$0.60 & 88.16 & 5.24\\
Fe\,I & 4882.14  & 3.41 & $-$1.48 & 19.96 & 5.42\\
Fe\,I & 4890.75  & 2.88 & $-$0.38 & 99.64 & 5.25\\
Fe\,I & 4891.49  & 2.85 & $-$0.14 & 108.09& 5.17 \\
Fe\,I & 4903.31  & 2.88 & $-$0.89 & 68.19 & 5.13 \\
Fe\,I & 4918.99  & 2.85 & $-$0.34 & 95.69 & 5.09 \\
Fe\,I & 4920.50  & 2.83 & 0.06    & 127.02& 5.34\\
Fe\,I & 4924.77  & 2.28 & $-$2.11 & 47.49 & 5.28\\
Fe\,I & 4938.81  & 2.87 & $-$1.08 & 60.97 & 5.17 \\
Fe\,I & 4939.69  & 0.86 & $-$3.25 & 94.18 & 5.60\\
Fe\,I & 4946.39  & 3.37 & $-$1.11 & 44.71 & 5.50\\
Fe\,I & 4950.11  & 3.41 & $-$1.50 & 11.14 & 5.14 \\
Fe\,I & 4966.09  & 3.33 & $-$0.79 & 45.59 & 5.15 \\
Fe\,I & 4973.10  & 3.96 & $-$0.69 & 16.85 & 5.16 \\
Fe\,I & 4985.25  & 3.93 & $-$0.44 & 38.36 & 5.35\\
Fe\,I & 4994.13  & 0.91 & $-$2.97 & 84.53 & 5.17 \\
Fe\,I & 5001.86  & 3.88 & $-$0.01 & 48.73 & 5.06 \\
Fe\,I & 5005.71  & 3.88 & $-$0.12 & 46.49 & 5.13 \\
Fe\,I & 5006.12  & 2.83 & $-$0.62 & 84.82 & 5.11 \\
Fe\,I & 5012.07  & 0.86 & $-$2.60 & 110.13& 5.29\\
Fe\,I & 5014.94  & 3.94 & $-$0.18 & 35.92 & 5.06 \\
Fe\,I & 5022.24  & 3.98 & $-$0.33 & 27.43 & 5.09 \\
Fe\,I & 5041.07  & 0.96 & $-$3.09 & 91.24 & 5.48\\
Fe\,I & 5041.76  & 1.48 & $-$2.20 & 96.95 & 5.35\\
Fe\,I & 5049.82  & 2.28 & $-$1.35 & 84.16 & 5.17 \\
Fe\,I & 5051.63  & 0.91 & $-$2.76 & 96.21 & 5.20\\
Fe\,I & 5060.08  & 0.00 & $-$5.43 & 28.38 & 5.55\\
Fe\,I & 5068.77  & 2.94 & $-$1.23 & 64.04 & 5.44\\
Fe\,I & 5074.75  & 4.22 & $-$0.20 & 37.15 & 5.42\\
Fe\,I & 5083.34  & 0.96 & $-$2.84 & 95.78 & 5.32\\
Fe\,I & 5090.77  & 4.25 & $-$0.40 & 25.57 & 5.42\\
Fe\,I & 5098.70  & 2.17 & $-$2.03 & 74.62 & 5.54\\
Fe\,I & 5166.28  & 0.00 & $-$4.12 & 91.58 & 5.34\\
Fe\,I & 5171.60  & 1.48 & $-$1.72 & 110.24& 5.13 \\
Fe\,I & 5191.45  & 3.04 & $-$0.55 & 96.15 & 5.50\\
Fe\,I & 5192.34  & 2.99 & $-$0.42 & 86.67 & 5.12 \\
Fe\,I & 5194.94  & 1.56 & $-$2.02 & 94.78 & 5.18\\
Fe\,I & 5198.71  & 2.22 & $-$2.09 & 50.59 & 5.23\\
Fe\,I & 5202.34  & 2.17 & $-$1.87 & 74.99 & 5.37\\
Fe\,I & 5215.18  & 3.26 & $-$0.86 & 46.03 & 5.13 \\
Fe\,I & 5216.27  & 1.61 & $-$2.08 & 89.99 & 5.19\\
Fe\,I & 5217.39  & 3.21 & $-$1.07 & 40.94 & 5.19\\
Fe\,I & 5225.52  & 0.11 & $-$4.75 & 49.26 & 5.35\\
Fe\,I & 5227.19  & 1.56 & $-$1.23 & 140.25& 5.33\\
Fe\,I & 5232.94  & 2.94 & $-$0.19 & 108.90& 5.28\\
Fe\,I & 5242.49  & 3.63 & $-$0.84 & 23.91 & 5.11 \\
Fe\,I & 5247.05  & 0.09 & $-$4.97 & 40.69 & 5.41\\
Fe\,I & 5250.21  & 0.12 & $-$4.90 & 41.11 & 5.38\\
Fe\,I & 5250.65  & 2.19 & $-$2.18 & 62.04 & 5.47\\
Fe\,I & 5253.46  & 3.28 & $-$1.58 & 14.46 & 5.17 \\
Fe\,I & 5254.96  & 0.11 & $-$4.76 & 58.82 & 5.51\\
Fe\,I & 5263.31  & 3.26 & $-$0.87 & 47.73 & 5.17 \\
Fe\,I & 5266.56  & 2.99 & $-$0.49 & 89.06 & 5.23\\
Fe\,I & 5281.79  & 3.04 & $-$1.02 & 64.35 & 5.34\\
Fe\,I & 5283.62  & 3.24 & $-$0.45 & 71.37 & 5.13 \\
Fe\,I & 5288.53  & 3.68 & $-$1.49 & 6.86  & 5.18\\
Fe\,I & 5302.30  & 3.28 & $-$0.73 & 55.06 & 5.17 \\
Fe\,I & 5307.36  & 1.61 & $-$2.91 & 44.59 & 5.21\\
Fe\,I & 5322.04  & 2.28 & $-$2.80 & 12.05 & 5.14 \\
Fe\,I & 5324.18  & 3.21 & $-$0.11 & 89.22 & 5.10 \\
Fe\,I & 5328.53  & 1.56 & $-$1.85 & 113.40& 5.38\\
Fe\,I & 5332.90  & 1.56 & $-$2.78 & 54.15 & 5.17 \\
Fe\,I & 5339.93  & 3.26 & $-$0.63 & 59.83 & 5.13 \\
Fe\,I & 5341.02  & 1.61 & $-$1.95 & 102.56& 5.31 \\
Fe\,I & 5364.87  & 4.44 & 0.22    & 36.26 & 5.22\\
Fe\,I & 5365.40  & 3.56 & $-$1.02 & 20.10 & 5.10 \\
Fe\,I & 5367.47  & 4.41 & 0.35    & 41.57 & 5.16 \\
Fe\,I & 5369.96  & 4.37 & 0.35    & 45.86 & 5.18\\
Fe\,I & 5371.49  & 0.96 & $-$1.64 & 149.17& 5.15 \\
Fe\,I & 5379.57  & 3.69 & $-$1.48 & 8.05  & 5.25\\
Fe\,I & 5383.37  & 4.31 & 0.50    & 56.71 & 5.15 \\
Fe\,I & 5389.48  & 4.41 & $-$0.41 & 15.25 & 5.32\\
Fe\,I & 5393.17  & 3.24 & $-$0.72 & 52.75 & 5.07 \\
Fe\,I & 5397.13  & 0.91 & $-$1.98 & 137.93& 5.22\\
Fe\,I & 5405.77  & 0.99 & $-$1.85 & 138.38& 5.19\\
Fe\,I & 5410.91  & 4.47 & 0.28    & 35.31 & 5.17 \\
Fe\,I & 5415.20  & 4.38 & 0.50    & 51.23 & 5.14 \\
Fe\,I & 5424.07  & 4.32 & 0.52    & 62.17 & 5.24\\
Fe\,I & 5429.70  & 0.96 & $-$1.88 & 142.47& 5.25\\
Fe\,I & 5434.52  & 1.01 & $-$2.13 & 125.98& 5.24\\
Fe\,I & 5476.56  & 4.10 & $-$0.28 & 29.58 & 5.20\\
Fe\,I & 5497.52  & 1.01 & $-$2.83 & 101.82& 5.43\\
Fe\,I & 5501.46  & 0.96 & $-$3.05 & 97.23 & 5.49\\
Fe\,I & 5506.78  & 0.99 & $-$2.79 & 97.65 & 5.28\\
Fe\,I & 5569.62  & 3.41 & $-$0.52 & 62.73 & 5.23\\
Fe\,I & 5572.84  & 3.39 & $-$0.28 & 73.61 & 5.16 \\
Fe\,I & 5576.09  & 3.43 & $-$1.00 & 46.26 & 5.45\\
Fe\,I & 5586.76  & 3.37 & $-$0.11 & 80.92 & 5.10 \\
Fe\,I & 5615.64  & 3.33 & 0.04    & 91.00 & 5.10 \\
Fe\,I & 5638.26  & 4.22 & $-$0.72 & 9.69  & 5.17 \\
Fe\,I & 5658.82  & 3.39 & $-$0.76 & 45.86 & 5.16 \\
Fe\,I & 5662.52  & 4.17 & $-$0.41 & 18.79 & 5.15 \\
Fe\,I & 5686.53  & 4.55 & $-$0.63 & 7.22  & 5.31\\
Fe\,I & 5701.54  & 2.56 & $-$2.22 & 28.16 & 5.31\\
Fe\,I & 5705.99  & 4.61 & $-$0.46 & 11.55 & 5.44\\
Fe\,I & 5753.12  & 4.25 & $-$0.62 & 10.33 & 5.15 \\
Fe\,I & 5775.08  & 4.22 & $-$1.08 & 4.84  & 5.20 \\
Fe\,I & 5816.37  & 4.55 & $-$0.68 & 7.95  & 5.40\\
Fe\,I & 5956.69  & 0.86 & $-$4.50 & 15.27 & 5.25\\
Fe\,I & 6003.01  & 3.88 & $-$1.10 & 12.02 & 5.26\\
Fe\,I & 6008.56  & 3.88 & $-$0.98 & 14.27 & 5.23\\
Fe\,I & 6065.48  & 2.61 & $-$1.53 & 63.27 & 5.27\\
Fe\,I & 6082.71  & 2.22 & $-$3.55 & 5.76  & 5.43\\
Fe\,I & 6136.62  & 2.45 & $-$1.40 & 82.01 & 5.28 \\
Fe\,I & 6136.99  & 2.19 & $-$2.93 & 21.77 & 5.43\\
Fe\,I & 6137.69  & 2.59 & $-$1.40 & 73.35 & 5.29\\
Fe\,I & 6151.62  & 2.17 & $-$3.37 & 7.51  & 5.32\\
Fe\,I & 6191.56  & 2.43 & $-$1.60 & 78.92 & 5.40\\
Fe\,I & 6200.31  & 2.61 & $-$2.44 & 15.98 & 5.25\\
Fe\,I & 6213.43  & 2.22 & $-$2.48 & 31.74 & 5.22\\
Fe\,I & 6219.28  & 2.19 & $-$2.45 & 42.12 & 5.35 \\
Fe\,I & 6230.72  & 2.56 & $-$1.28 & 82.42 & 5.29\\
Fe\,I & 6232.64  & 3.65 & $-$1.24 & 14.86 & 5.23\\
Fe\,I & 6240.65  & 2.22 & $-$3.17 & 10.03 & 5.30\\
Fe\,I & 6246.32  & 3.60 & $-$0.77 & 35.37 & 5.19\\
Fe\,I & 6252.56  & 2.40 & $-$1.69 & 68.79 & 5.27\\
Fe\,I & 6254.26  & 2.28 & $-$2.43 & 42.23 & 5.42\\
Fe\,I & 6265.13  & 2.17 & $-$2.54 & 41.24 & 5.39\\
Fe\,I & 6280.62  & 0.86 & $-$4.39 & 37.28 & 5.61\\
Fe\,I & 6297.79  & 2.22 & $-$2.64 & 40.51 & 5.54\\
Fe\,I & 6301.50  & 3.65 & $-$0.71 & 32.89 & 5.14 \\
Fe\,I & 6322.69  & 2.59 & $-$2.43 & 20.04 & 5.33 \\
Fe\,I & 6335.33  & 2.19 & $-$2.18 & 49.24 & 5.19\\
Fe\,I & 6336.82  & 3.68 & $-$0.85 & 28.68 & 5.23\\
Fe\,I & 6355.03  & 2.84 & $-$2.42 & 17.82 & 5.56\\
Fe\,I & 6393.60  & 2.43 & $-$1.62 & 77.85 & 5.38\\
Fe\,I & 6400.00  & 3.60 & $-$0.27 & 61.38 & 5.14 \\
Fe\,I & 6411.65  & 3.65 & $-$0.59 & 41.79 & 5.18\\
Fe\,I & 6421.35  & 2.28 & $-$2.01 & 60.57 & 5.29\\
Fe\,I & 6430.85  & 2.17 & $-$1.95 & 69.95 & 5.27\\
Fe\,I & 6494.98  & 2.40 & $-$1.27 & 93.30 & 5.27\\
Fe\,I & 6498.94  & 0.96 & $-$4.69 & 10.04 & 5.32\\
Fe\,I & 6592.91  & 2.73 & $-$1.60 & 63.56 & 5.46\\
Fe\,I & 6593.87  & 2.43 & $-$2.42 & 32.66 & 5.41\\
Fe\,I & 6609.11  & 2.56 & $-$2.69 & 14.89 & 5.40\\
Fe\,I & 6663.44  & 2.42 & $-$2.48 & 29.97 & 5.40\\
Fe\,I & 6677.98  & 2.69 & $-$1.47 & 72.76 & 5.43\\
Fe\,I & 6750.15  & 2.42 & $-$2.62 & 29.00 & 5.52\\
Fe\,I & 7511.02  & 4.17 & 0.12    & 59.11 & 5.32 \\
Fe\,II & 4178.86 & 2.58 & $-$2.51 & 73.89 & 5.25\\
Fe\,II & 4489.19 & 2.83 & $-$2.96 & 44.41 & 5.39\\
Fe\,II & 4491.41 & 2.86 & $-$2.71 & 45.59 & 5.20 \\
Fe\,II & 4508.28 & 2.86 & $-$2.44 & 65.85 & 5.30\\
Fe\,II & 4515.34 & 2.84 & $-$2.60 & 59.81 & 5.33\\
Fe\,II & 4520.22 & 2.81 & $-$2.65 & 59.38 & 5.33\\
Fe\,II & 4555.89 & 2.83 & $-$2.40 & 65.27 & 5.21\\
Fe\,II & 4576.34 & 2.84 & $-$2.95 & 44.31 & 5.39\\
Fe\,II & 4582.84 & 2.84 & $-$3.18 & 28.74 & 5.32\\
Fe\,II & 4583.84 & 2.81 & $-$1.93 & 90.08 & 5.23\\
Fe\,II & 4620.52 & 2.83 & $-$3.21 & 27.55 & 5.31\\
Fe\,II & 4731.44 & 2.89 & $-$3.10 & 29.79 & 5.31\\
Fe\,II & 4993.35 & 2.81 & $-$3.62 & 15.17 & 5.35 \\
Fe\,II & 5197.58 & 3.23 & $-$2.22 & 52.27 & 5.21\\
Fe\,II & 5234.63 & 3.22 & $-$2.18 & 59.49 & 5.29\\
Fe\,II & 5264.81 & 3.23 & $-$3.13 & 16.03 & 5.35\\
Fe\,II & 5276.00 & 3.20 & $-$2.01 & 69.79 & 5.27\\
Fe\,II & 5284.08 & 2.89 & $-$3.11 & 29.93 & 5.30\\
Fe\,II & 5325.55 & 3.22 & $-$3.16 & 13.23 & 5.27\\
Fe\,II & 5414.07 & 3.22 & $-$3.58 & 7.36  & 5.40\\
Fe\,II & 5534.83 & 3.25 & $-$2.75 & 27.47 & 5.28\\
Fe\,II & 6247.54 & 3.89 & $-$2.30 & 17.73 & 5.29\\
Fe\,II & 6432.68 & 2.89 & $-$3.57 & 15.35 & 5.34\\
Co\,I & 3842.05  & 0.92 & $-$0.77 & syn   & 3.00 \\
Co\,I & 3873.12  & 0.43 & $-$0.66 & syn   & 2.70 \\
Co\,I & 3995.31  & 0.92 & $-$0.22 & syn   & 2.40 \\
Co\,I & 4110.53  & 1.05 & $-$1.08 & syn   & 2.80 \\
Co\,I & 4118.77  & 1.05 & $-$0.49 & syn   & 2.80 \\
Co\,I & 4121.31  & 0.92 & $-$0.32 & syn   & 2.74 \\
Ni\,I & 3452.89  & 0.11 & $-$0.90 & 179.17& 4.25\\
Ni\,I & 3472.54  & 0.11 & $-$0.79 & 142.46& 3.71\\
Ni\,I & 3483.78  & 0.27 & $-$1.11 & 124.67& 3.89\\
Ni\,I & 3492.96  & 0.11 & $-$0.24 & 206.92& 3.80\\
Ni\,I & 3500.85  & 0.16 & $-$1.27 & 114.66& 3.68\\
Ni\,I & 3519.76  & 0.27 & $-$1.44 & 121.24& 4.12 \\
Ni\,I & 3524.54  & 0.03 & 0.01    & 249.37& 3.68\\
Ni\,I & 3566.37  & 0.42 & $-$0.24 & 157.55& 3.69\\
Ni\,I & 3597.70  & 0.21 & $-$1.10 & 121.02& 3.66\\
Ni\,I & 3783.53  & 0.42 & $-$1.40 & 137.46& 4.37\\
Ni\,I & 3858.30  & 0.42 & $-$0.96 & 158.50& 4.24\\
Ni\,I & 4604.99  & 3.48 & $-$0.24 & 16.72 & 3.87\\
Ni\,I & 4648.65  & 3.42 & $-$0.09 & 23.80 & 3.84\\
Ni\,I & 4686.21  & 3.59 & $-$0.59 & 6.42  & 3.88\\
Ni\,I & 4714.42  & 3.38 & 0.25    & 74.28 & 4.41\\
Ni\,I & 4855.41  & 3.54 & 0.00    & 25.69 & 3.92\\
Ni\,I & 4904.41  & 3.54 & $-$0.17 & 24.72 & 4.07 \\
Ni\,I & 4980.17  & 3.60 & 0.07    & 25.42 & 3.91\\
Ni\,I & 5035.36  & 3.63 & 0.29    & 27.40 & 3.77\\
Ni\,I & 5080.53  & 3.65 & 0.32    & 34.05 & 3.90\\
Ni\,I & 5081.11  & 3.84 & 0.30    & 24.42 & 3.93\\
Ni\,I & 5084.09  & 3.68 & 0.03    & 18.97 & 3.87\\
Ni\,I & 5137.07  & 1.68 & $-$1.94 & 45.01 & 4.06 \\
Ni\,I & 5476.90  & 1.83 & $-$0.78 & 90.31 & 3.85\\
Ni\,I & 5578.72  & 1.68 & $-$2.83 & 17.02 & 4.33 \\
Ni\,I & 5754.66  & 1.93 & $-$2.22 & 17.42 & 4.03 \\
Ni\,I & 6108.12  & 1.68 & $-$2.60 & 14.53 & 3.99\\
Ni\,I & 6643.63  & 1.68 & $-$2.22 & 33.93 & 4.05 \\
Ni\,I & 6767.77  & 1.83 & $-$2.14 & 29.49 & 4.05 \\
Zn\,I & 4722.15  & 4.03 & $-$0.34 & 38.70 & 2.71\\
Zn\,I & 4810.53  & 4.08 & $-$0.14 & 34.98 & 2.48\\
Sr\,II & 3464.46 & 3.04 & 0.53    & syn   & 1.00 \\
Sr\,II & 4077.71 & 0.17 & 0.00    & syn   & 1.05 \\
Sr\,II & 4161.79 &$-$0.6& 2.94    & syn   & 1.23 \\
Y\,II & 3611.04  & 0.13 & 0.11    & syn   & 0.27 \\
Y\,II & 3710.29  & 0.18 & 0.46    & syn   & 0.15 \\
Y\,II & 3774.34  & 0.13 & 0.21    & syn   & 0.60 \\
Y\,II & 4398.01  & 0.13 & $-$1.00 & syn   & 0.37 \\
Y\,II & 4682.33  & 0.41 & $-$1.51 & syn   & 0.40 \\
Y\,II & 4854.87  & 0.99 & $-$0.38 & syn   & 0.25 \\
Y\,II & 4883.68  & 1.08 & 0.07    & syn   & 0.40 \\
Y\,II & 4900.11  & 1.03 & $-$0.09 & syn   & 0.20 \\
Y\,II & 4982.13  & 1.03 & $-$1.29 & syn   & 0.19 \\
Y\,II & 5087.42  & 1.08 & $-$0.17 & syn   & 0.30 \\
Y\,II & 5119.11  & 0.99 & $-$1.36 & syn   & 0.77 \\
Y\,II & 5200.41  & 0.99 & $-$0.57 & syn   & 0.22 \\
Y\,II & 5205.73  & 1.03 & $-$0.34 & syn   & 0.30 \\
Y\,II & 5289.82  & 1.03 & $-$1.85 & syn   & 0.55 \\
Y\,II & 5320.78  & 1.08 & $-$1.95 & syn   & 0.45 \\
Y\,II & 5473.39  & 1.73 & $-$1.02 & syn   & 0.65 \\
Zr\,II & 3457.55 & 0.56 & $-$0.53 & 58.55 & 0.88\\
Zr\,II & 3458.94 & 0.96 & $-$0.52 & 40.59 & 0.94 \\
Zr\,II & 3499.57 & 0.41 & $-$0.81 & syn   & 0.78 \\
Zr\,II & 3505.67 & 0.16 & $-$0.36 & 89.45 & 1.09 \\
Zr\,II & 3536.94 & 0.36 & $-$1.31 & 37.35 & 0.94\\
Zr\,II & 3573.08 & 0.32 & $-$1.04 & 62.57 & 1.17 \\
Zr\,II & 3607.38 & 1.27 & $-$0.64 & syn   & 1.07 \\
Zr\,II & 3630.03 & 0.36 & $-$1.11 & 46.35 & 0.91\\
Zr\,II & 3698.15 & 1.01 & 0.094   & 65.37 & 0.80\\
Zr\,II & 3714.79 & 0.53 & $-$0.93 & 52.21 & 0.97\\
Zr\,II & 3766.82 & 0.41 & $-$0.81 & 61.81 & 0.90\\
Zr\,II & 3998.95 & 0.56 & $-$0.67 & 66.57 & 0.99\\
Zr\,II & 4050.33 & 0.71 & $-$1.00 & syn   & 0.80 \\
Zr\,II & 4071.10 & 1.00 & $-$1.60 & syn   & 1.03 \\
Zr\,II & 4090.51 & 0.76 & $-$1.10 & 41.15 & 1.15 \\
Zr\,II & 4149.20 & 0.80 & $-$0.03 & syn   & 0.90 \\
Zr\,II & 4161.21 & 0.71 & $-$0.72 & syn   & 1.10 \\
Zr\,II & 4208.99 & 0.71 & $-$0.46 & syn   & 0.80 \\
Zr\,II & 4317.32 & 0.71 & $-$1.38 & syn   & 1.00 \\
Zr\,II & 5112.28 & 1.66 & $-$0.59 & syn   & 0.71 \\
Ru\,I & 3728.03  & 0.00 & 0.27    & syn   & 0.34 \\
Ru\,I & 3742.28  & 0.34 & $-$0.18 & syn   & 0.52 \\
Ru\,I & 3798.90  & 0.15 & $-$0.04 & syn   & 0.61 \\
Ru\,I & 3799.35  & 0.00 & 0.02    & syn   & 0.70 \\
Rh\,I & 3692.36  & 0.00 & 0.17    & syn   & $-$0.40  \\
Pd\,I & 3404.58  & 0.81 & 0.32    & syn   & $-$0.25 \\
Pd\,I & 3460.74  & 0.81 & $-$0.42 & syn   & $-$0.02 \\
Pd\,I & 3516.94  & 0.94 & $-$0.24 & syn   & 0.14  \\
Sn\,I & 3801.02  & 1.07 & 0.74    & syn   & $-$0.24  \\
Ba\,II & 4130.65 & 2.72 & 0.68    & syn   &  0.94  \\
Ba\,II & 4554.03 & 0.00 & 0.14    & syn   &  0.82  \\
Ba\,II & 4934.10 & 0.00 & $-$0.16 & syn   &  1.14  \\
Ba\,II & 5853.69 & 0.60 & $-$0.91 & syn   &  0.83  \\
Ba\,II & 6141.73 & 0.70 & $-$0.08 & syn   &  0.90  \\
Ba\,II & 6496.91 & 0.60 & $-$0.38 & syn   &  1.05  \\
La\,II & 4086.71 & 0.00 & $-$0.07 & syn   & $-$0.11  \\
La\,II & 4322.51 & 0.17 & $-$0.93 & syn   & $-$0.10  \\
La\,II & 4333.75 & 0.17 & $-$0.06 & syn   &    0.30  \\
La\,II & 4429.91 & 0.23 & $-$0.35 & syn   & $-$0.05  \\
La\,II & 4526.12 & 0.77 & $-$0.59 & syn   & $-$0.07  \\
La\,II & 4558.46 & 0.32 & $-$0.97 & syn   & $-$0.10  \\
La\,II & 4574.88 & 0.17 & $-$1.08 & syn   & $-$0.11  \\
La\,II & 4662.51 & 0.00 & $-$1.24 & syn   & $-$0.08  \\
La\,II & 4748.73 & 0.93 & $-$0.54 & syn   & $-$0.13  \\
La\,II & 4804.04 & 0.23 & $-$1.49 & syn   &  0.12  \\
La\,II & 4809.00 & 0.24 & $-$1.4  & syn   & $-$0.10  \\
La\,II & 4921.78 & 0.24 & $-$0.45 & syn   & $-$0.10  \\
La\,II & 4986.83 & 0.17 & $-$1.30 & syn   & $-$0.04  \\
La\,II & 5114.56 & 0.23 & $-$1.03 & syn   & $-$0.05  \\
La\,II & 5122.99 & 0.32 & $-$0.85 & syn   & $-$0.10  \\
La\,II & 5259.38 & 0.17 & $-$1.95 & syn   & $-$0.10  \\
La\,II & 6262.29 & 0.40 & $-$1.22 & syn   & $-$0.05  \\
La\,II & 6390.48 & 0.32 & $-$1.41 & syn   &  0.00  \\
Ce\,II & 3534.04 & 0.52 & $-$0.14 & 28.19 & 0.34\\
Ce\,II & 3539.08 & 0.32 & $-$0.27 & 24.03 & 0.13 \\
Ce\,II & 3659.23 & 0.17 & $-$0.67 & 27.19 & 0.37 \\
Ce\,II & 3912.42 & 0.29 & $-$0.25 & 33.48 & 0.19 \\
Ce\,II & 3993.82 & 0.91 & 0.29    & 54.11 & 0.77\\
Ce\,II & 3999.24 & 0.29 & 0.06    & 56.88 & 0.33\\
Ce\,II & 4053.50 & 0.00 & $-$0.61 & 32.33 & 0.16 \\
Ce\,II & 4068.84 & 0.70 & $-$0.17 & 24.99 & 0.39\\
Ce\,II & 4083.22 & 0.70 & 0.27    & 50.98 & 0.47\\
Ce\,II & 4118.14 & 0.70 & 0.13    & 48.01 & 0.54\\
Ce\,II & 4120.83 & 0.32 & $-$0.37 & 40.01 & 0.44\\
Ce\,II & 4127.36 & 0.68 & 0.31    & 35.64 & 0.10 \\
Ce\,II & 4137.65 & 0.52 & 0.40    & 52.19 & 0.13 \\
Ce\,II & 4145.00 & 0.70 & 0.10    & 34.80 & 0.31\\
Ce\,II & 4222.60 & 0.12 & $-$0.15 & 46.90 & 0.10 \\
Ce\,II & 4349.77 & 0.53 & $-$0.73 & 24.91 & 0.71\\
Ce\,II & 4364.65 & 0.50 & $-$0.17 & 32.83 & 0.28\\
Ce\,II & 4486.91 & 0.29 & $-$0.18 & 45.49 & 0.27\\
Ce\,II & 4523.08 & 0.52 & $-$0.08 & 45.18 & 0.43\\
Ce\,II & 4560.28 & 0.91 & 0.18    & 58.83 & 0.88 \\
Ce\,II & 4560.96 & 0.68 & $-$0.26 & 19.75 & 0.27\\
Ce\,II & 4562.36 & 0.48 & 0.21    & 55.57 & 0.27 \\
Ce\,II & 4572.28 & 0.68 & 0.22    & 47.69 & 0.36\\
Ce\,II & 4582.50 & 0.70 & $-$0.35 & 22.22 & 0.44\\
Ce\,II & 4593.93 & 0.70 & 0.07    & 43.42 & 0.45\\
Ce\,II & 4628.16 & 0.52 & 0.14    & 47.77 & 0.24 \\
Ce\,II & 5187.46 & 1.21 & 0.17    & 13.56 & 0.21\\
Ce\,II & 5274.23 & 1.04 & 0.13    & 19.94 & 0.25\\
Ce\,II & 5330.56 & 0.87 & $-$0.40 & 9.99  & 0.22\\
Ce\,II & 6043.37 & 1.20 & $-$0.48 & 5.17  & 0.35\\
Pr\,II & 4062.81 & 0.42 & 0.33    & syn   & $-$0.54  \\
Pr\,II & 4222.95 & 0.06 & 0.27    & syn   & $-$0.65  \\
Pr\,II & 4468.26 & 0.22 & $-$0.23 & syn   & $-$0.60  \\
Pr\,II & 4496.47 & 0.06 & $-$0.27 & syn   & $-$0.53  \\
Pr\,II & 4510.15 & 0.42 & $-$0.02 & syn   & $-$0.63  \\
Pr\,II & 5173.91 & 0.97 & 0.38    & syn   & $-$0.64  \\
Pr\,II & 5206.55 & 0.22 & $-$0.05 & syn   & $-$0.30  \\
Pr\,II & 5219.04 & 0.79 & $-$0.05 & syn   & $-$0.68  \\
Pr\,II & 5322.76 & 0.48 & $-$0.32 & syn   & $-$0.90  \\
Nd\,II & 3887.9  & 0.04 & $-$0.78 & 39.7  & 0.19 \\
Nd\,II & 3900.22 & 0.47 & 0.10    & 47.54 & 0.11 \\
Nd\,II & 3990.10 & 0.47 & 0.13    & 51.66 & 0.14 \\
Nd\,II & 4004.00 & 0.06 & $-$0.57 & 33.14 & 0.00 \\
Nd\,II & 4011.06 & 0.47 & $-$0.76 & 9.85  & 0.00 \\
Nd\,II & 4012.70 & 0.00 & $-$0.60 & 33.04 & $-$0.05 \\
Nd\,II & 4023.00 & 0.56 & 0.04    & 36.50 & 0.04 \\
Nd\,II & 4041.06 & 0.47 & $-$0.53 & 21.61 & 0.18 \\
Nd\,II & 4043.59 & 0.32 & $-$0.71 & 19.07 & 0.11 \\
Nd\,II & 4051.14 & 0.38 & $-$0.30 & 30.00 & 0.03 \\
Nd\,II & 4061.08 & 0.47 & 0.55    & 63.44 & $-$0.05 \\
Nd\,II & 4069.26 & 0.06 & $-$0.57 & 36.24 & 0.05 \\
Nd\,II & 4109.4  & 0.06 & 0.35    &  85.5 & 0.48 \\
Nd\,II & 4133.35 & 0.32 & $-$0.49 & 37.25 & 0.29\\
Nd\,II & 4232.37 & 0.06 & $-$0.47 & 44.66 & 0.09 \\
Nd\,II & 4368.63 & 0.06 & $-$0.81 & 30.21 & 0.13 \\
Nd\,II & 4446.38 & 0.20 & $-$0.35 & 39.27 & 0.01 \\
Nd\,II & 4451.98 & 0.00 & $-$1.10 & 22.79 & 0.17 \\
Nd\,II & 4463.0  & 0.00 & 0.40    & 68.2  & 0.42  \\
Nd\,II & 4465.59 & 0.18 & $-$1.10 & 13.55 & 0.11 \\
Nd\,II & 4501.79 & 0.18 & $-$0.69 & 50.32 & 0.32 \\
Nd\,II & 4542.60 & 0.74 & $-$0.28 & 15.61 & 0.02 \\
Nd\,II & 4563.22 & 0.18 & $-$0.88 & 23.10 & 0.16 \\
Nd\,II & 4645.76 & 0.56 & $-$0.76 & 10.69 & 0.09 \\
Nd\,II & 4706.54 & 0.00 & $-$0.71 & 38.61 & 0.10 \\
Nd\,II & 4709.69 & 0.18 & $-$0.97 & 18.85 & 0.13 \\
Nd\,II & 4715.62 & 0.18 & $-$0.90 & 19.9  & 0.11 \\
Nd\,II & 4825.48 & 0.18 & $-$0.42 & 43.12 & 0.08 \\
Nd\,II & 4902.59 & 0.18 & $-$1.34 & 14.65 & 0.21 \\
Nd\,II & 4914.38 & 0.38 & $-$0.70 & 17.39 & 0.03 \\
Nd\,II & 4959.12 & 0.06 & $-$0.80 & 30.05 & 0.06 \\
Nd\,II & 4987.19 & 0.18 & $-$0.79 & 12.87 & 0.39 \\
Nd\,II & 5063.69 & 0.38 & $-$0.62 & 7.25  & 0.21 \\
Nd\,II & 5092.79 & 0.38 & $-$0.61 & 20.67 & 0.02 \\
Nd\,II & 5130.59 & 1.30 & 0.45    & 24.17 & 0.14 \\
Nd\,II & 5212.36 & 0.20 & $-$0.96 & 19.09 & 0.11 \\
Nd\,II & 5234.19 & 0.55 & $-$0.51 & 20.51 & 0.11 \\
Nd\,II & 5249.58 & 0.98 & 0.20    & 24.76 & 0.01 \\
Nd\,II & 5255.51 & 0.20 & $-$0.67 & 30.27 & 0.07 \\
Nd\,II & 5273.36 & 0.82 & $-$0.18 & 54.91 & 0.36 \\
Nd\,II & 5311.45 & 0.99 & $-$0.42 & 6.62  & $-$0.03 \\
Nd\,II & 5319.81 & 0.55 & $-$0.14 & 35.38 & 0.05 \\
Nd\,II & 5357.02 & 0.55 & $-$0.28 & 8.15  & 0.25 \\
Nd\,II & 5371.92 & 0.55 & 0.00    & 12.12 & 0.11 \\
Nd\,II & 5485.70 & 1.26 & $-$0.12 & 11.86 & 0.26\\
Nd\,II & 5740.86 & 1.16 & $-$0.53 & 3.92  & 0.02 \\
Sm\,II & 3979.20 & 0.54 & $-$0.47 & 13.62 & $-$0.30\\
Sm\,II & 4068.32 & 0.43 & $-$0.76 & 9.52  & $-$0.33 \\
Sm\,II & 4206.12 & 0.38 & $-$0.72 & 8.72  & $-$0.49\\
Sm\,II & 4220.66 & 0.54 & $-$0.44 & 9.58  & $-$0.53\\
Sm\,II & 4318.93 & 0.28 & $-$0.25 & 24.17 & $-$0.57\\
Sm\,II & 4424.34 & 0.48 & 0.14    & 33.91 & $-$0.51\\
Sm\,II & 4499.48 & 0.25 & $-$0.87 & 9.24  & $-$0.50\\
Sm\,II & 4511.83 & 0.18 & $-$0.82 & syn   & $-$0.38  \\
Sm\,II & 4519.63 & 0.54 & $-$0.35 & syn   & $-$0.50  \\
Sm\,II & 4537.94 & 0.49 & $-$0.48 & syn   & $-$0.25  \\
Sm\,II & 4554.44 & 0.10 & $-$1.25 & syn   & $-$0.40  \\
Sm\,II & 4566.20 & 0.33 & $-$0.59 & 12.47 & $-$0.53\\
Sm\,II & 4591.81 & 0.18 & $-$1.12 & syn   & $-$0.33  \\
Sm\,II & 4595.28 & 0.49 & $-$0.5  & syn   & $-$0.36  \\
Sm\,II & 4604.17 & 0.04 & $-$1.39 & syn   & $-$0.43  \\
Sm\,II & 4615.44 & 0.54 & $-$0.69 & 8.69  & $-$0.37\\
Sm\,II & 4642.23 & 0.38 & $-$0.46 & 23.20 & $-$0.29\\
Sm\,II & 4669.39 & 0.10 & $-$0.60 & 24.15 & $-$0.46\\
Sm\,II & 4719.84 & 0.04 & $-$1.24 & syn   & $-$0.24  \\
Eu\,II & 3724.93 & 0.00 & $-$0.09 & syn   & $-$1.09  \\
Eu\,II & 4129.72 & 0.22 & 0.22    & syn   & $-$1.13  \\
Eu\,II & 6437.64 & 1.32 & $-$0.32 & syn   & $-$1.05  \\
Eu\,II & 6645.10 & 1.25 & 0.12    & syn   & $-$1.10  \\
Gd\,II & 3481.80 & 0.49 & 0.12    & 19.23 & $-$0.51\\
Gd\,II & 3549.36 & 0.24 & 0.29    & 38.68 & $-$0.51\\
Gd\,II & 4037.89 & 0.56 & $-$0.42 & 7.94  & $-$0.45\\
Gd\,II & 4085.56 & 0.73 & $-$0.01 & 12.72 & $-$0.43\\
Gd\,II & 4251.73 & 0.38 & $-$0.22 & 20.02 & $-$0.41\\
Tb\,II & 3658.89 & 0.13 & $-$0.01 & syn   & $-$1.28  \\
Tb\,II & 3702.85 & 0.13 & 0.44    & syn   & $-$1.15  \\
Tb\,II & 3899.19 & 0.37 & 0.33    & syn   & $-$1.50  \\
Tb\,II & 4002.57 & 0.64 & 0.10    & syn   & $-$1.00  \\
Dy\,II & 3506.81 & 0.10 & $-$0.60 & 21.42 & $-$0.39\\
Dy\,II & 3536.02 & 0.54 & 0.53    & 43.75 & $-$0.48\\
Dy\,II & 3694.81 & 0.10 & $-$0.11 & syn   & $-$0.65  \\
Dy\,II & 3983.65 & 0.54 & $-$0.31 & 33.04 & 0.01 \\
Dy\,II & 3996.69 & 0.59 & $-$0.26 & syn   & $-$0.35  \\
Dy\,II & 4050.57 & 0.59 & $-$0.47 & syn   & $-$0.19  \\
Dy\,II & 4073.12 & 0.54 & $-$0.32 & syn   & $-$0.41  \\
Dy\,II & 4077.97 & 0.10 & $-$0.04 & 54.72 & $-$0.38\\
Ho\,II & 3456.01 & 0.00 & 0.76    & syn   & $-$1.31  \\
Ho\,II & 3474.27 & 0.08 & 0.28    & syn   & $-$1.50  \\
Ho\,II & 3484.83 & 0.08 & 0.28    & syn   & $-$1.10  \\
Ho\,II & 3810.71 & 0.00 & 0.19    & syn   & $-$0.67  \\
Ho\,II & 3890.97 & 0.08 & 0.46    & syn   & $-$0.80  \\
Ho\,II & 4045.45 & 0.00 & $-$0.05 & syn   & $-$1.10  \\
Er\,II & 3633.54 & 0.00 & $-$0.53 & 21.45 & $-$0.69\\
Er\,II & 3692.65 & 0.06 & 0.28    & 64.55 & $-$0.54 \\
Er\,II & 3729.52 & 0.00 & $-$0.59 & 22.08 & $-$0.67\\
Er\,II & 3906.31 & 0.00 & 0.12    & 66.09 & $-$0.47\\
Tm\,II & 3462.20 & 0.00 & 0.03    & syn   & $-$1.26  \\
Tm\,II & 3701.36 & 0.00 & $-$0.54 & syn   & $-$1.30  \\
Tm\,II & 3761.91 & 0.00 & $-$0.43 & syn   & $-$1.55  \\
Tm\,II & 3996.51 & 0.00 & $-$1.20 & syn   & $-$1.50  \\
Lu\,II & 3472.48 & 1.54 & $-$0.19 & syn   & $-$1.20 \\
Hf\,II & 3918.09 & 0.45 & $-$1.14 & syn   & $-$0.70  \\
Hf\,II & 4093.15 & 0.45 & $-$1.15 & syn   & $-$0.60  \\
Os\,I &  4260.85 & 0.00 & $-$1.44 & syn   & $-$0.25  \\
Os\,I &  4420.46 & 0.00 & $-$1.53 & syn   & $-$0.01  \\
Ir\,I &  3513.65 & 0.00 & $-$1.26 & syn   & $-$0.25  \\
Pb\,I &  4057.81 & 1.32 & $-$0.17 & syn   &  1.21  \\
Th\,II &  4019.12  & 0.00 & $-$0.23 & syn & $-$1.70  \\
\enddata
\end{deluxetable}

\acknowledgements We thank Chris Sneden for providing an up-to-date version of his neutron-capture element linelists. M.G. and M.C. acknowledge support from the MIT UROP program. 
A.F. is supported by NSF-CAREER grant AST-1255160. C.A. acknowledges funding from the Alexander von Humboldt 
Foundation.
This work benefited from support by the National Science Foundation under Grant No. PHY-1430152; Physics Frontier Center / JINA Center for the Evolution of the Elements (JINA-CEE).
APJ is supported by NASA through Hubble Fellowship grant HST-HF2-51393.001 awarded by the Space Telescope Science Institute, which is operated by the Association of Universities for Research in Astronomy, Inc., for NASA, under contract NAS5-26555.
This work made extensive use of NASA's Astrophysics Data System Bibliographic Services and the python libraries 
\texttt{numpy} \citep{numpy}, 
\texttt{scipy} \citep{scipy}, 
\texttt{matplotlib} \citep{matplotlib},
and \texttt{astropy} \citep{astropy}
\facilities{Magellan:Clay (MIKE), ESO/NTT, SOAR 4.1m}
\software{MOOG \citep{Sneden73}, MIKE Carnegie Python Pipeline \citep{Kelson03}, IRAF \citep{tody86}, binary c/nucsyn
(\citealt{Izzard2004,Izzard2006,Izzard2009})}.

\vspace{2cm}
\appendix

We present additional investigations into one CEMP-$s$ and three \textquotedblleft CEMP-$r/s$" stars. We show that their neutron-capture abundances signatures are qualitatively different from that of \thestar\ , and cannot be explained with the same origin scenario, i.e., they are not $r+s$ stars. Our findings regarding these stars are in line with results from previous studies (e.g., \citealt{Hampel16}). Only in the case of \thestar\ (as described in the paper) is our model of an initial (independent) r-process enhancement followed by a s-process binary pollution scenario able to reproduce the observed abundance pattern. 


\section{One CEMP-$s$ and three \textquotedblleft CEMP-$r/s$" stars from the literature} 

We investigate the abundance patterns of one CEMP-$s$ star, CS 22881–036, and three \textquotedblleft CEMP-$r/s$" stars, CS 22948–027, CS 29497–030, and LP $625–44$, by applying the same procedure as for \thestar\ (see Section 4.2 for more details) to assess whether principally different origin scenarios are required. In the following we provide details on the results of each star.

\subsection{CS 22881-036}
\begin{figure} [ht!]
\includegraphics[width = 7.cm, angle=270]{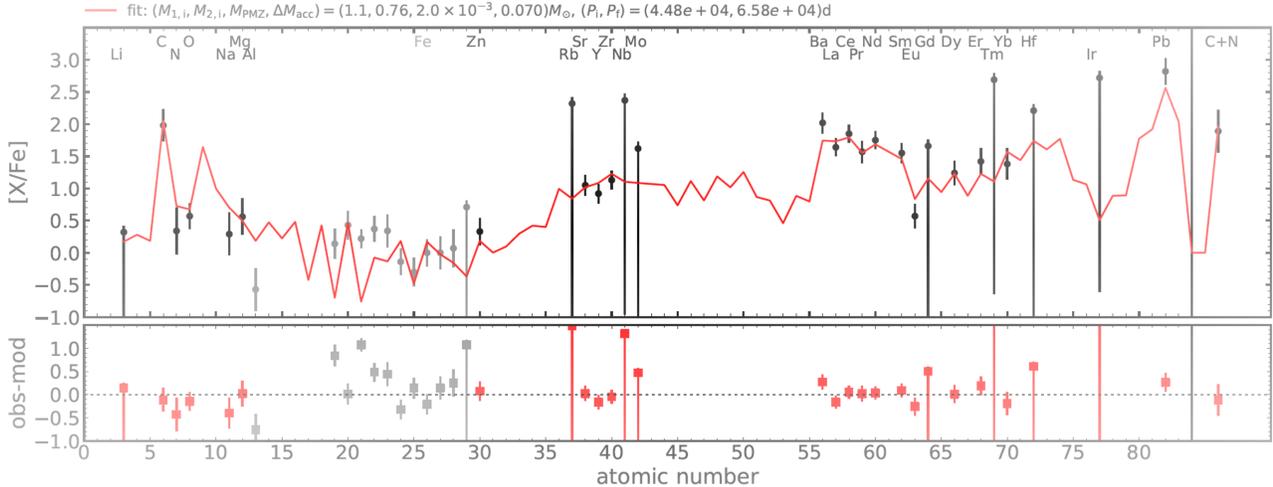}
\caption{Best fit (red line) to the observed abundances of CS 22881-036 (filled circles), with residuals. Elements with very large error bars are upper limits. The parameters of the fit are shown on top. No pre-enrichment in $r$-process elements was adopted in this model.}
\label{CS22881}
\end{figure}

We used abundances from \citet{roederer2014}. The best fit to the observed abundances of CS 22881-036 is shown in Figure~\ref{CS22881}. To reproduce the abundances, it was not necessary to add any pre-enrichment in $r$-process elements. $\mbox{[Ba/Eu]}= +1.4$ is sufficiently high, and the Eu abundance sufficiently low ($\mbox{[Eu/Fe]} = +0.58$) to indicate that a pure $s$-process from an AGB star companion produces enough (i.e., the observed) Eu. Adding any $r$-process elements actually decreases the goodness of the fit to the abundances, suggesting that it is indeed a pure $s$-process-enhanced star. 
 
\subsection{CS 22948-027}
\begin{figure} [ht!]

\includegraphics[width = 7.2cm,angle=270]{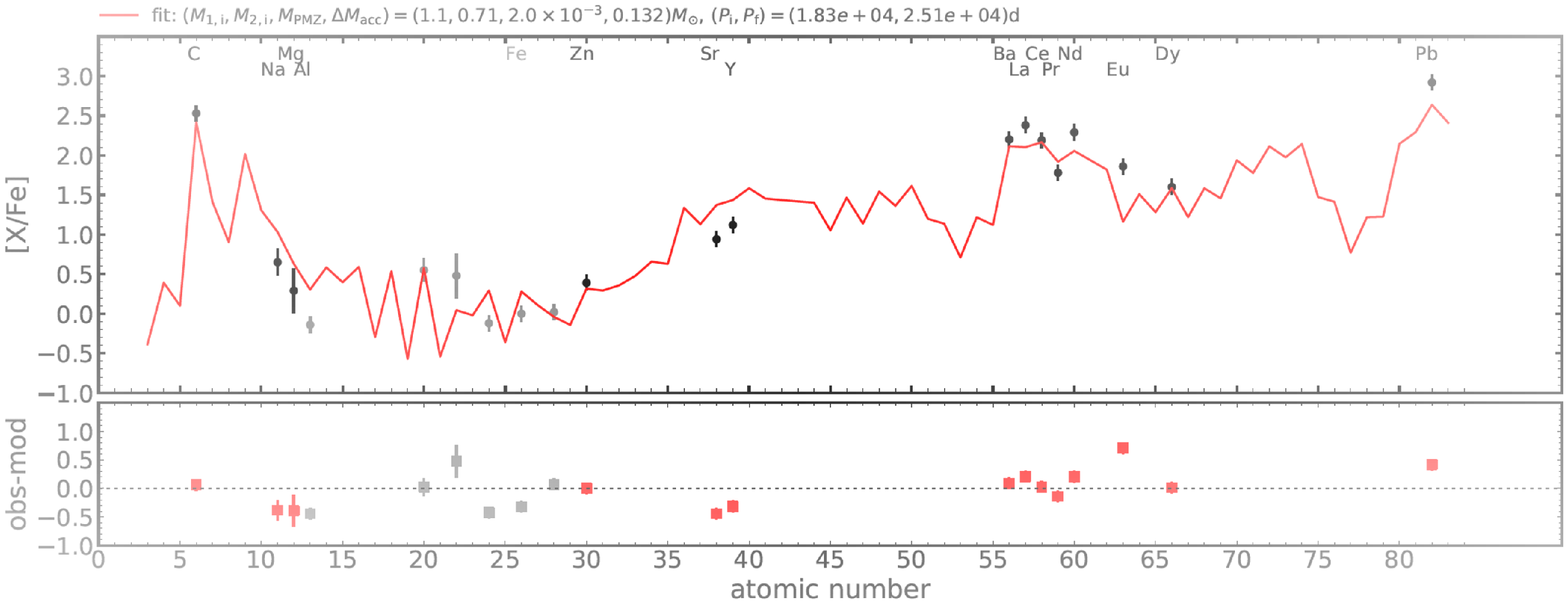}\\
\includegraphics[width = 7.2cm, angle=270]{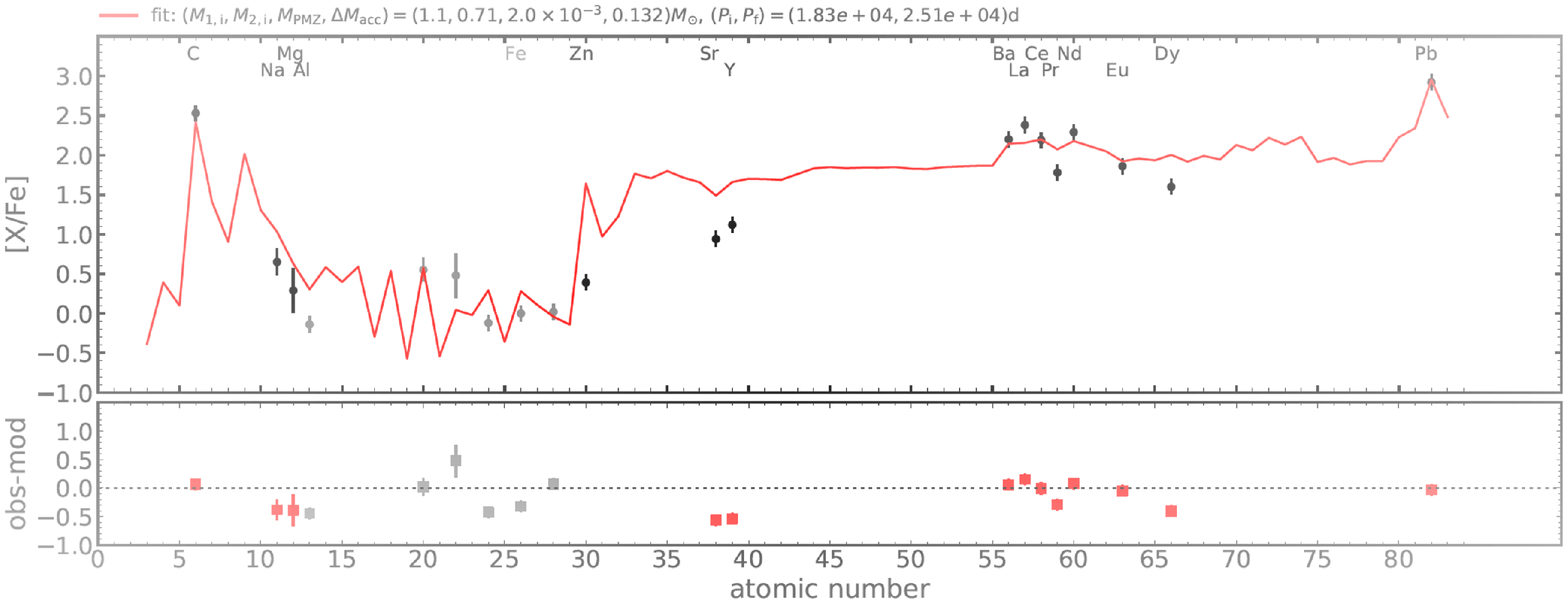}
\caption{Best fit (red line) to the observed abundances of CS22948-027 (filled circles) with residuals. The parameters of the fit are shown on top. The top double panel shows results with no pre-enrichment in $r$-process elements (that is, the initial abundance of all elements from zinc to thorium is solar-scaled down to metallicity Z$=10^{-4}$). 
The lower double panel shows the results assuming a pre-enrichment in $r$-elements with a contribution that results in the observed Eu abundance of $\mbox{[Eu/Fe]} = +1.86$. The other neutron-capture elements are accordingly rescaled using the $r$-pattern of \citet{burris2000}.
} \label{CS22948}
\end{figure}

We used abundances from \citet{barbury2005}. The best fits to the observed abundances of CS 22948-027, without and with the contribution of a pre-enrichment in $r$-process elements, are shown in Figure~\ref{CS22948}. The case with no pre-enrichment (top panel) does not provide a satisfactory fit to the elemental abundances. For example, the model abundance of Eu is too low, by 0.7\,dex, and Pb is under-produced by $\sim$0.5\,dex. 

Adding an initial abundance of Eu to reproduce the observed value of $\mbox{[Eu/Fe]} = +1.86$ by definition provides the required Eu abundance but otherwise produces a rather poor result for all other heavy elements (bottom panel). The abundances of e.g., Ba, La, Ce, are dominated by the $s$-process elements contributed by the AGB star, so any initial amounts of these elements added are essentially washed out, resulting in no significantly different fit in that region. In addition, the abundance of Zn is over-estimated by more than 1\,dex, and Sr and Y are over-estimated by almost as much, resulting in a poor overall fit. 

Overall, the enhancements in neutron-capture elements in this star are large but not extreme. The biggest problem with fitting the abundances is the element-to-element ratios. This is common with models trying to reproduce \textquotedblleft CEMP-r/s" stars (e.g., Abate et al. 2015a). The model cannot simultaneously reproduce the ratios [C/hs], [C/ls], [hs/ls] and [Pb/hs], and in
particular, any large ratios [hs/ls]$>$ 1 and [Pb/hs]$>$ 1 (where hs and ls refer to light and heavy $s$-process elements, respectively). Therefore, the best fit is found by a model that passes somewhere \textquotedblleft in the middle", thus over-estimating the elements in the first peak (here Sr and Y) and under-estimating some of the heavier elements. This kind of element distribution is thus much better reproduced by an $i$-process model (i.e., \citealt{Hampel16}).

\subsection{CS 29497-030}
\begin{figure} [ht!]
\includegraphics[width = 7.3cm,angle=270]{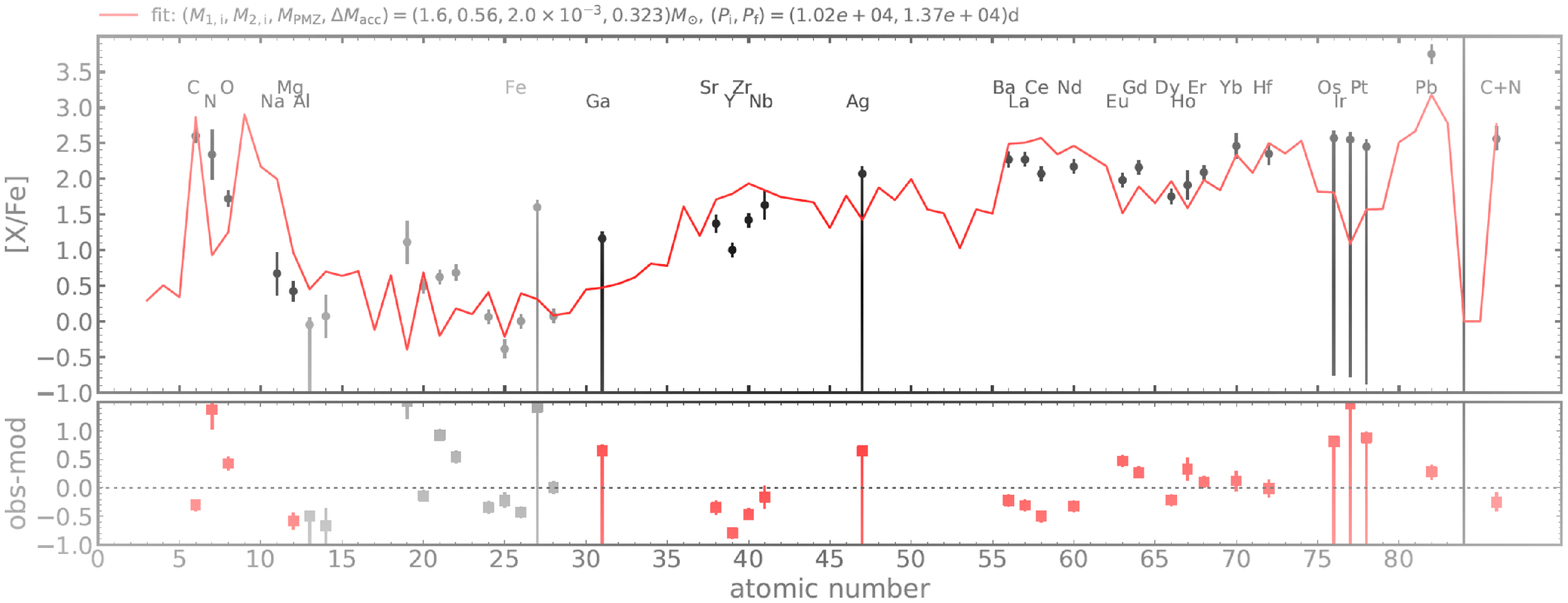}
\includegraphics[width = 7.cm,angle=270]{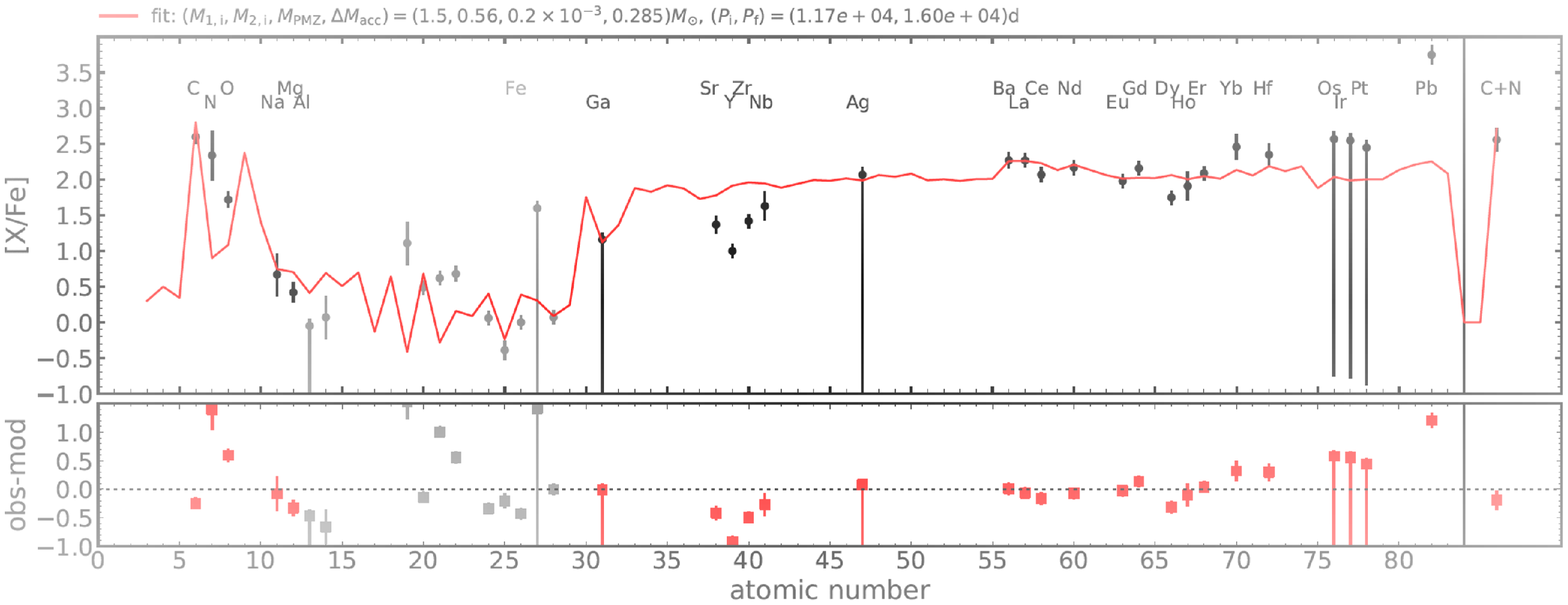}
\caption{ Best fit (red line) to the observed abundances of CS29497-030 (filled circles) with residuals. Elements with very large error bars are upper limits.  The parameters of the fit are shown on top. The top double panel shows results with no pre-enrichment in $r$-elements. 
The lower double panel shows the pattern assuming a pre-enrichment in $r$-elements with a contribution that results in the observed Eu abundance of $\mbox{[Eu/Fe]} = +1.98$. The other neutron-capture elements are accordingly rescaled using the $r$-pattern of  \citet{burris2000}.
}
\label{CS29497}
\end{figure}

We used abundances from \citet{ivans05}. The best fits to the observed abundances of CS 29497-030, without and with the contribution of a pre-enrichment in $r$-process elements, are shown in Figure~\ref{CS29497}, respectively. The overall case is similar as for the that of CS 22948-027. Since CS 29497-030 has a very large over-abundance of Pb, combined with the high abundances of neutron-capture elements between Ba and Pt, it is essentially impossible to find model parameters which simultaneously reproduce the various element-to-element ratios. 

Adding an initial $r$-process enhancement to the model somewhat improves the fit to the heavier elements, except in the case of Pb. At these AGB masses (1.5\,M$_{\odot}$), the abundances of Na, Mg, and Pb are quite sensitive to the mass of the partial-mixing zone ($M_{\mathrm{PMZ}}$). In the pure $s$-process model, the only way to produce large abundances of the heavies elements (Eu, Gd, Ho, Yb, Pb) is through a rather large $M_{\mathrm{PMZ}}$. Consequently, the abundances of Na and Mg will be large and thus highly over-estimated. Instead, in the pre-enriched model, the abundances of most of these heavy elements are provided through the initial $r$-process enhancement. Therefore, it is possible to find a good fit to those elements already with a relatively small $M_{\mathrm{PMZ}}$. The model also better reproduces the abundances of Na and Mg, but at the expense of a worse fit to Pb. The pre-enriched model thus yields an overall better fit, although still not a good fit, suggesting that an $i$-process model might be more suitable.

\subsection{LP 625-44}

\begin{figure} [ht!]
\includegraphics[width = 7.5cm,angle=270]{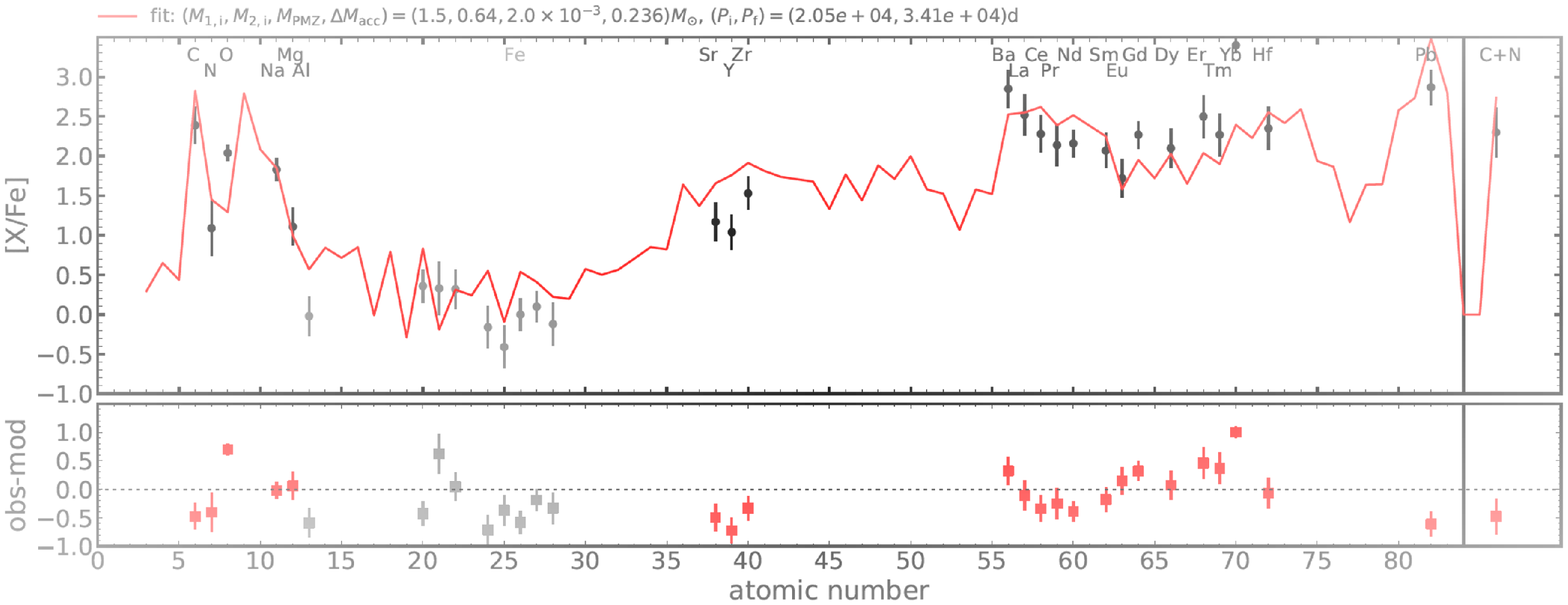}
\includegraphics[width = 7.5cm,angle=270]{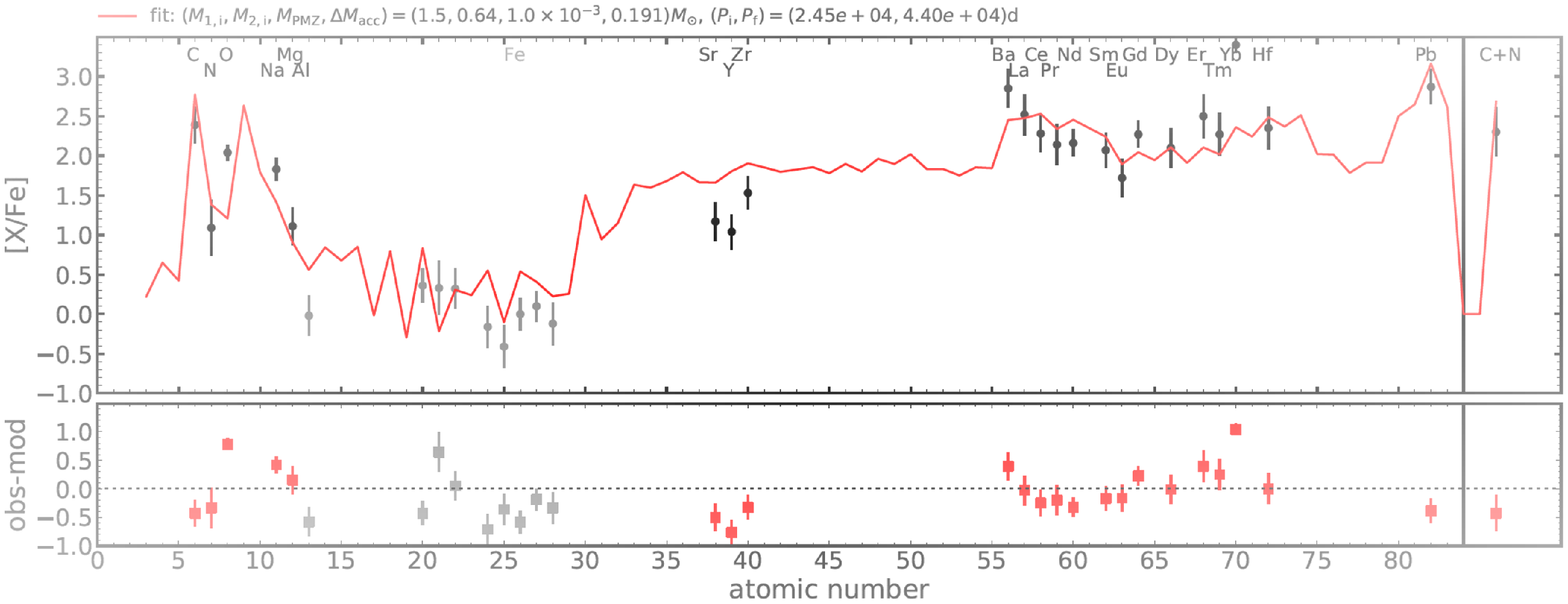}
\caption{Best fit (red line) to the observed abundances of LP625-44 (filled circles) with residuals. Elements with very large error bars are upper limits.  The parameters of the fit are shown on top. The top double panel shows results with no pre-enrichment in $r$-elements. 
The lower double panel shows the pattern assuming a pre-enrichment in $r$-elements with a contribution that results in the observed Eu abundance of $\mbox{[Eu/Fe]} = +1.72$. The other neutron-capture elements are accordingly rescaled using the $r$-pattern of  \citet{burris2000}}
\label{LP625}
\end{figure}

We used abundances from \citet{aoki2002}. The best fits to the observed abundances of LP 625-44, without and with the contribution of a pre-enrichment in $r$-elements, are shown in Figure~\ref{LP625}. As for the stars discussed above, reproducing element-to-element ratios is challenging. Although some of the elements are highly enriched (e.g., $\mbox{[Eu/Fe]} = +1.72$), their abundance could be entirely produced by an $s$-process in an AGB star (the exception would be Yb, which is anomalously enhanced, possibly pointing to an observational problem). However, an AGB star that produces the observed amounts of e.g., Ba, Eu,and Pb,  would also produce large abundances of C, Sr, Y, and Zr. Consequently, these elements are all over-estimated by the model by 0.5 to 1\,dex. 

The best fit does not significantly improve using an $r$-process pre-enriched set of abundances because these initial abundances are partially washed out by any of the newly-produced $s$-process material. Interestingly, adding an initial enrichment implies that the secondary star can accrete significantly less material (20\% less
accreted mass) to reach essentially the same enhanced abundances. Regardless, as is the case for CS22948–027 and CS29497–030, the \textquotedblleft best fit" is not matching the data at all, thus suggesting a different origin scenario for the star.

\end{document}